\documentclass[a4paper,11pt]{article}
\usepackage{algorithm}
\usepackage{algpseudocode}
\usepackage{classeJBArxive}
\usepackage{tabularx}
\usepackage{caption}

\newcommand{\degC}{^{\,\circ}C}
\newcommand*\unit[1]{\bigl[\, \mathsf{#1} \,\bigr]}
\definecolor{dkgreen}{rgb}{0,0.52,0.14}
\definecolor{purple2}{rgb}{255,0,255}

\title{\vspace{-1.5cm}
Estimation of thermal properties and boundary heat transfer coefficient of the ground with a Bayesian technique \\	
\vspace{4pt}}

\author{Zhanat Karashbayeva \textsuperscript{a, b}$^{\ast}$, Julien Berger\textsuperscript{a}, Helcio R.B. Orlande\textsuperscript{c}, Marie-Hélène Azam\textsuperscript{d}}
\date{\vspace{-0.8cm}}
\begin{document}

\maketitle

\begin{center}
\small
\textsuperscript{a} Laboratoire des Sciences de l’Ingénieur pour l’Environnement (LaSIE), UMR 7356 CNRS, La Rochelle Université, CNRS, 17000, La Rochelle, France\\
\textsuperscript{b} Astana IT University, Astana, Kazakhstan \\
\textsuperscript{c} POLI/COPPE, Mechanical Engineering Department, Federal University of Rio de Janeiro \\
\textsuperscript{d} Université de Strasbourg, INSA Strasbourg, CNRS, ICube Laboratory UMR 7357, Equipe GCE, Strasbourg, France \\
$^{\ast}$corresponding author, e-mail address : zhanat.karashbaeva@astanait.edu.kz\\

\end{center}

\begin{abstract}
Urbanization is the key contributor for climate change. Increasing urbanization rate causes an urban heat island (UHI) effect, which strongly depends on the short- and long-wave radiation balance heat flux between the surfaces. In order to calculate accurately this heat flux, it is required to assess the surface temperature which depends on the knowledge of the thermal properties and the surface heat transfer coefficients in the heat transfer problem. The aim of this paper is to estimate the thermal properties of the ground and the time varying surface heat transfer coefficient by solving an inverse problem. The  \textsc{Dufort}--\textsc{Frankel} scheme is applied for solving the unsteady heat transfer problem. For the inverse problem, a Markov chain Monte Carlo method is used to estimate the posterior probability density function of unknown parameters within the Bayesian framework of statistics, by applying the Metropolis-Hastings algorithm for random sample generation. Actual temperature measurements available at different ground depths were used for the solution of the inverse problem. Different time discretizations were examined for the transient heat transfer coefficient at the ground surface, which then involved different prior distributions. Results of different case studies show that the estimated values of the unknown parameters were in accordance with literature values. Moreover, with the present solution of the inverse problem the temperature residuals were smaller than those obtained by using literature values for the unknowns.   


\end{abstract}

\section{Introduction}

As recently reported, the global urbanization rate is above $50 \%$ and it is predicted to increase in the future. Urbanization is becoming one of the main causes of  global warming. Current studies indicates that the urban heat island (UHI) phenomenon is rapidly increasing in the areas with higher urbanization rates \cite{ZHANG2025111877}. The heat flux from ground is one of the main reason for the development of the UHI effect. Ground heat flux is the most important component of the surface energy balance \cite{AZAM201822, MUSY2015213, DONG2025112311, ELIZONDOMARTINEZ2020100967}. Therefore, accurate computational simulations of heat transfer processes in the ground are of great importance for the prediction of UHI effects \cite{AZAM2018728,COLOMBERT2011125, REES20071478}. For the mathematical modeling, thermal properties of the ground such as thermal conductivity and volumetric heat capacity are key parameters for energy balances \cite{HU20161,KERAVECBALBOT2025105995}. Consequently, it is very important to accurately estimate these thermal property values of the ground. As reported in \cite{NWAKAIRE2020102476}, pavements and grounds are crucial for effective mitigation strategies to to ensure lower surface temperature and reduce urban heat island effects. Moreover, the boundary conditions of the physical model plays an important role, especially at the top of the ground \cite{MIRZAEI2015200}. Energy balance at the surface of the ground is composed of different heat flux components. According to \cite{HERB2008327} the convection heat flux, represented in terms of the surface heat transfer coefficient, can have a significant impact on the surface temperature. 

There is an important need to determine the thermophysical properties of the pavement materials as well as the surface heat transfer coefficients at the interface between the ground and the outside ambient air \cite{SREEDHAR201678}. Some studies available in the literature were dedicated to the direct measurement of parameters appearing in models for ground heat transfer, but with destructive and/or expensive methods \cite{PALACIOS201932, KRASTEV20103847, SAUER1995161, JAKKAREDDY2018144}. Inverse problems can also be used for the identification of model parameters and functions with \textit{in-situ} ground measurements. Inverse problems based on actual \textit{in-situ} experimental data are complicated, since their solutions are influenced by the measurement and model errors, due to their ill-posed character. Consequently, most of the available works devoted to the estimation of the thermal properties of the ground are based on simulated measurements \cite{Huntul2020102, Santos2006161} or on the laboratory experiments \cite{ALPAR2024124727, UKRAINCZYK20095675, Yang20051}, which need validation with real field experiments. Moreover, most of the works consider the surface heat transfer coefficient as constant \cite{Berger_bayesian_2023}. However it varies according to time, as it is influenced by the velocity of the wind at the top of the surface. In \cite{CHANTASIRIWAN19994275} a time-dependent heat transfer coefficient is estimated, but using simulated measurements. Analysis of the state-of-the-art show that the simultaneous estimation of the thermal properties of the ground and of the time varying surface heat transfer coefficient  with real \textit{in-situ} measurement data is still an open question.

The objective of this work is to apply a Bayesian approach for simultaneously estimating the thermal conductivity, volumetric heat capacity and time varying surface heat transfer coefficient of a ground using actual temperature measurements at different depths below the surface \cite{COHARD2018675}. Several works are available in the literature on the use of this method for the solution of inverse problems using Bayesian approach in heat transfer \cite{Ozisik, Orlande_2010}. However, in these works the Bayesian approach was applied using simulated measurements. 

This paper is structured as follows: Section \ref{Methodology} presents the description of the physical model. Section \ref{ssec:Parest} gives information about the inverse problem method and its
algorithm. In Section \ref{ssec:casestudy}, a case study of surface heat transfer in ground is investigated where the thermal conductivity, volumetric heat capacity and time varying surface heat transfer coefficient are retrieved. Last, in Section~\ref{sec:urban_scale}, a comparison at the urban scale is performed between a standard simulation and one using the estimated parameters. 

\section{Methodology}
\label{Methodology}

\subsection{Description of Physical Model} 

The physical problem considers one-dimensional transient heat conduction in grounds defined by the
spatial domain $\Omega_{\,x} = [0, L]$, where $L \ \unit{m}$ is the ground depth and $\Omega_{\,t} = [0, t_{\,f}]$ is the time domain, where $t_{\,f} \ \unit{s}$ is the time horizon for investigating the phenomena. The physical problem can
be formulated as \cite{BERGER2021101849}:
\begin{equation} \label{heatequation} 
   C \frac {\partial T}{\partial t} = \frac{\partial }{\partial x} \left(\kappa    \frac{\partial T}{\partial x} \right), 
\end{equation}
where $C = \rho c_{\,\rho} \ \unit{J \cdot m^{-3}\cdot K^{-1}}$ is the volumetric heat capacity,
where $\rho \ \unit{kg \cdot m^{-3}}$ is the density, $c_{\,\rho} \ \unit{J \cdot kg^{-1}\cdot K^{-1}}$ is the specific heat, $\kappa \ \unit{W\cdot m^{-1} \cdot K^{-1}}$ is the thermal conductivity, and $T \ \unit{K}$ is the temperature of the ground depending on spatial $x$ and time $t$ variables. 

The boundary condition at the surface of the ground defines the balance between diffusive, radiation and convection fluxes. It is expressed as:

\begin{equation} \label{BCL}
    \kappa \frac{\partial T}{\partial x} = h (\,  t \,) \left(T - T_{\,\infty}(\,  t \,)\right) - q_{\,\infty} \left( \, t \, \right) , \qquad \forall t\in \Omega_{\,t}\,, \qquad x = 0\,, 
\end{equation}
where $T_{\,\infty} \ \unit{K}$ is the temperature of the air and $q_{\,\infty} \ \unit{W\cdot m^{-2}}$ is the net radiation flux, which are time-dependent functions due to climate variations. Both vary according to the climate. $h \ \unit{W\cdot m^{-2} \cdot K^{-1}}$ is the surface heat transfer coefficient. 

\begin{figure}[htbp]
  \centering
  \includegraphics[width = 0.75\textwidth]{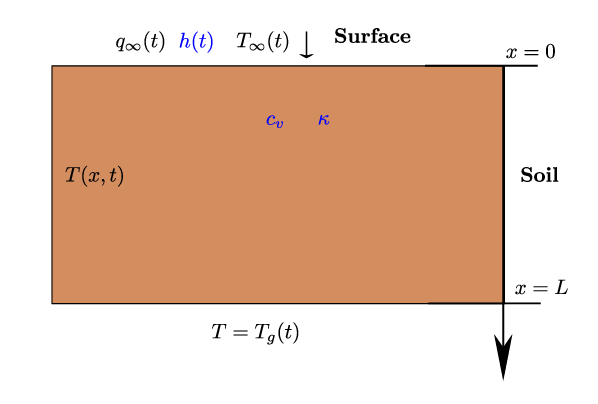}
  \caption{Illustration of the physical model. }
  \label{fig:illustration_problem}
\end{figure}

At the depth of $x = L$, a \textsc{Dirichlet} boundary condition is set:
\begin{equation} \label{BCR}
    T \egal T_{g} (t) , \qquad \forall t\in \Omega_{\,t}\,, \qquad x \egal L \,. 
\end{equation}
The initial condition is:
\begin{equation} \label{IC}
    T \egal T_{\,in}(x), \qquad \forall x\in \Omega_{\,x}\,, \qquad t \egal 0 \,. 
\end{equation}
The physical model is illustrated in Figure~\ref{fig:illustration_problem}. 

Equations \ref{heatequation} to \ref{IC} define the direct or forward problem, so that the distribution of temperature $T(x,t)$ can be computed when the physical properties, initial and boundary conditions are known.

\subsection{Dimensionless formulation}

This section presents the dimensionless form of equations \eqref{heatequation}--\eqref{IC}. The following dimensionless variables are defined:
\begin{align*}
    t^* \egal \frac{t}{t_{\,ref}} \,, \qquad x^* \egal \frac{x}{L} \,,
\end{align*}

\begin{align*}
    U \egal \frac{T}{T_{\,ref}} \,, \qquad U_{\,g} \egal \frac{T_{\,g}}{T_{\,ref}}, \qquad U_{\,\infty} \egal \frac{T_{\,\infty}}{T_{\,ref}} \qquad U_{\,in} = \frac{T_{\,in}}{T_{\,ref}} \,,
\end{align*}
and the dimensionless coefficients:
\begin{align*}
    k \egal \frac{\kappa}{\kappa_{ref}} \,, \quad
     C^* \egal \frac{C}{C_{ref}} \,, \quad h^* \egal \frac{h}{h_{ref}} \,, \quad Fo \egal \frac{\kappa_{ref} \cdot t_{\,ref} }{c_{\,v_{ref}} \cdot L^2} \,, \quad Bi^L \egal \frac{h_{\,ref} \cdot L}{\kappa_{ref}} \,, \quad q_{\,\infty}^* \egal \frac{q_{\,\infty} \cdot L}{\kappa_{ref} \cdot T_{ref}} \,.
\end{align*}
Thus, the heat equation \eqref{heatequation} is re-written as:
\begin{equation} \label{heateqDimless} 
    C^*  \frac {\partial U}{\partial t^*} \egal Fo \ \frac{\partial }{\partial x^*} \left( k \frac{\partial U}{\partial x^*} \right) \,,
\end{equation}
with the \textsc{Robin} and \textsc{Dirichlet}–type boundary conditions:
\begin{equation}
\label{RobinDimless} 
    k\frac{\partial U }{\partial x^*} = h^* (t^*) Bi^L \left(U - U_{\,\infty}(t^*)  \right) - q_{\,\infty}^* \,, \qquad \forall t^*\in \Omega_{\,t^*}\,, \qquad x^* = 0\,, 
\end{equation}

\begin{equation}
     \label{DirichletDimless}
    U(1,t^*) \egal U_{g}(t^*) \,, \qquad \forall t^*\in \Omega_{\,t^*}\,, \qquad x^* \egal 1 \,, 
\end{equation}
and the initial condition:
\begin{equation} \label{ICDimless}
    U \egal U_{\,in}(x^*), \qquad \forall x^* \in \Omega_{\,x^*}\,, \qquad t^* \egal 0 \,.
\end{equation}

The governing equation is solved using \textsc{Du} \textsc{Fort}--\textsc{Frankel} scheme~\cite{Gasparin_2017b}. All the implementation is realized in the \textsc{Matlab} environment.

\section{Inverse Problem via Bayesian inference}
\label{ssec:Parest}


The inverse problem aims at the estimation of the thermal conductivity ($\kappa$), surface heat transfer coefficient ($h(t)$) and volumetric heat capacity ($C$). Since the ground is assumed homogeneous, the thermal conductivity and volumetric heat capacity coefficients are considered as constants.
The surface heat transfer coefficient is defined as:
\begin{equation}
       \label{h(t)}
    h (t) \egal  \sum_{i=1}^{N_t} h_i \ \eta_i(t), 
\end{equation}
where $\eta_i$ is the piecewise basis function:
\begin{align*}
   \eta_i(t) \egal  \begin{cases}
      \, 1\,, \qquad t_i \leq t \leq t_{i+1} \,,  \quad \forall i \, \in \, \bigl\{\, 1\,,\,\ldots\,,\,N_t\,\bigr\}\\
      \, 0 \,, \qquad $otherwise$. 
    \end{cases}  
\end{align*}

The unknown parameters $\textbf{P} \in \Omega_P$ are denoted by:
\begin{align*} 
   \textbf{P} \stackrel{\text{def}}{: =} (p_1 \,, p_2 \,, p_3 \,, p_4 \,, ... \,, p_{N}) \equiv (\kappa \,, c_{\,v} \,, h_1 \,, h_2 \,, ... \,, h_{N_t}) \,,
\end{align*}
where $N = 2 + N_t$ is the number of unknowns.
For the solution of the  inverse problem, transient temperature measurements taken at different depths of the ground are given. The temperature measurements are denoted as $\textbf{Y}$. The measurement errors are supposed as uncorrelated Gaussian random variables, with zero means and known variance. The measurement errors are additive and independent of the unknown parameters $\textbf{P}$.

\subsection{Sensitivity Coefficients}

It is important to analyze the sensitivity coefficients before attempting to solve the parameter estimation problem, in order to evaluate the identifiability of the unknowns. The sensitivity coefficient is defined by:
\begin{align*}
X_{p_{ij}} = \frac{\partial T}{\partial p_j}\,\Bigr|_{\,x \egal x_{\,j}}  \,, \qquad \forall (\,i\,,\,j\,) \, \in \, \bigl\{\, 1\,,\,\ldots\,,\,M\,\bigr\} \, \times \, \bigl\{\, 1\,,\,\ldots\,,\,N\,\bigr\} \,,
\end{align*} 
where $i$ stands for the measurement positions. Different approaches exist for the computation of sensitivity coefficients \cite{Orlande_2011}. In our case, the central finite-difference approximation is used,
\begin{align*} 
    X_{p_{ij}}  \approx \frac{T(x_i\,, t\,, p_1 \,, p_2 \,, ...\,, p_j + \Delta p_j\,, ...\,, p_N) - T(x_i\,, t\,, p_1\,,p_2\,, ...\,,p_j - \Delta p_j\,, ...\,,p_N)}{2\, \Delta p_j }  \,,
\end{align*}
where the parameter mesh is taken as $\Delta \, p_j = 10^{-2} \, p_j$ in our computations.

\subsection{Bayesian estimation}

The Markov Chain Monte Carlo (MCMC) is applied to estimate the posterior distribution of the unknown parameters within the Bayesian framework of statistics \cite{Walter_1982, Kaipio2006}. According to Bayes’ theorem, we have \cite{beck_1977, Kaipio2006, calvetti2007, Lanzetta2011}:

\begin{align*}
\label{posterior}
     \pi_{posterior} \textbf{(P)} = \pi \textbf{(P|Y)} = \frac{\pi \textbf{(P)} \pi \textbf{(Y|P)}}{\pi \textbf{(Y)}} \,,
\end{align*}
where $\pi_{posterior}\textbf{(P)} $ is the posterior probability density, $\pi \textbf{(P)}$ is the prior density, $\pi \textbf{(Y|P)}$ is the likelihood function, and $\pi \textbf{(Y)}$ is the marginal probability density of the measurements. The latter is generally difficult to estimate and
not required from a practical point of view to determine the unknown parameters. Given that, Bayes' theorem is simplified to:

\begin{align*}
     \pi \textbf{(P|Y)}  \propto \pi \textbf{(P)} \pi \textbf{(Y|P)} \,.
\end{align*}
With the above hypotheses regarding the measurement errors, the likelihood is expressed as \cite{Kaipio2006, kaipio2011bayesian, Orlande_2012}:

\begin{align*}
    \pi \textbf{(Y|P)} = (2 \, \pi)^{-D / 2} | \textbf{W} |^{-1 / 2} \exp \left ( - \frac{1}{2} [\textbf{Y} - \textbf{T(P)}]^T \textbf{W}^{-1} [\textbf{Y} - \textbf{T(P)}]   \right )  
\end{align*}
where $D$ is the total number of measurements, and $\textbf{T(P)}$ is the solution of the direct problem \eqref{heatequation} - \eqref{IC}, obtained at the each sensor position and given parameters $\textbf{P}$. 

Two different prior probability densities are examined in this work. One of them is a Gaussian distribution in the domain $\Omega_P$:

\begin{align*}
     \pi \textbf{(P)} = (2 \, \pi)^{-N / 2} | \textbf{V} |^{-1 / 2} \exp \left \{ - \frac{1}{2} [\textbf{P} -  \boldsymbol{\mu}]^T \textbf{V}^{-1} [\textbf{P} - \boldsymbol{\mu}]   \right \} \,,
\end{align*}
where $\boldsymbol{\mu}$ and $\textbf{V}$ are the known mean and covariance matrix of the prior, respectively. As the parameters are assumed independent, the prior covariance matrix is diagonal, with elements given by the variances, $\sigma_i^2$, that is,

\begin{align*}
   [\textbf{V}_{i,j}] \egal  \begin{cases}
      \sigma_i^2\,, \qquad i = j \,, \qquad  \forall (\,i\,,\,j\,) \, \in \, \bigl\{\, 1\,,\,\ldots\,,\,N\,\bigr\}^{\,2}  \,, \\
      \, 0 \,, \qquad $otherwise$. 
    \end{cases}  
\end{align*}
The other prior density is the Gaussian smoothness prior, which is popular for priors in inverse problems for estimating parameters that approximate local values of functions \cite{Kaipio2006}. In this work the Gaussian smoothness prior is only applied for the surface heat transfer coefficient $h(t)$, Equation \eqref{h(t)}. It is given in the following form:

\begin{align*}
     \pi \textbf{(P)} = (2 \, \pi)^{-N_t / 2} \gamma^{N_t / 2} | \textbf{Z}^{-1} |^{-1 / 2} \exp \left \{ - \frac{1}{2} \gamma [\textbf{P} -  \Tilde{\textbf{P}}]^T \textbf{Z} [\textbf{P} - \Tilde{\textbf{P}}]\right \} \,,
\end{align*}
where $\textbf{Z} = \textbf{D}^T \textbf{D}$ and  $\textbf{D}$ is a $(\textbf{I}-1)\times \textbf{I}$ first-order difference matrix given in next form: 

\begin{align*}
\textbf{D} = \begin{bmatrix}
-1   &    1 & 0   & ... &   0\\
0 &  -1  & 1 & ... & 0 \\
... & ... & ... & ... & ... \\
0 & ... & 0 & -1 & 1
\end{bmatrix} \,,
\end{align*}
and the parameter $\gamma$ is treated in this work as a hyperparameter, that is, an unknown parameter of the model of the posterior distribution. In this work we consider the hyperparameter density in the form of the Rayleigh distribution \cite{Kaipio2006}:  

\begin{equation} \label{Rayleigh_distr}
     \pi (\gamma)  =\frac{\gamma}{\gamma_0^2} \exp \left[ -  \frac{1}{2} \left( \frac{\gamma}{\gamma_0}\right)^2\right] \,,
\end{equation}
where $\gamma_0$ is the scale parameter and it is defined by user.

The \textsc{Metrolpolis-Hastings} algorithm is used in this work to explore the posterior distribution. The implementation of this algorithm starts with the selection of a proposal distribution $q(\textbf{P}^*|\textbf{P}^{(k)})$, which is used to draw a new candidate sample $\textbf{P}^*$ given the current sample $\textbf{P}^{(k)}$ of the Markov chain (Step 2). Here a uniform proposal is used:

\begin{align*}
 q(\textbf{P}^*|\textbf{P}^{(k)}) = \textbf{P}^{(k)} + \omega \, \textbf{U}([-1\,,1]),
\end{align*}
where $\omega$ is the maximum step that can be taken from $\textbf{P(k)}$ to generate $\textbf{P}^*$. Then, the solution of the direct problem, Equations \eqref{heatequation} - \eqref{IC}, is computed
given the sampled parameter $\textbf{P}^*$ (Step 3). After that, the posterior distribution is evaluated
together with the acceptance factor defined by:

\begin{equation} \label{alpha}
    \alpha (\textbf{P}^*|\textbf{P}^{(k)}) = \min \left[1, \frac{\pi_{posterior}(\textbf{P}^*)\, q(\textbf{P}^{(k)}|\textbf{P}^*)}{\pi_{posterior}(\textbf{P}^{(t)}) \, q(\textbf{P}^*|\textbf{P}^{(k)})} \right].
\end{equation}

A random value U uniformly distributed on $[0, 1]$ is then generated (Step 5). At Step 6, if $U \leq \alpha (\textbf{P}^*|\textbf{P}^{(k)})$ then
the candidate parameter is retained $\textbf{P}^{k+1} = \textbf{P}^{(*)}$. Otherwise, set $\textbf{P}^{k+1} = \textbf{P}^{(k)}$. A sequence of sampled parameters is then obtained $\{\textbf{P}^{(1)} \,, \textbf{P}^{(2)} \,, ... \,, \textbf{P}^{(N_s)}\}$.
The \textsc{Metrolpolis-Hastings} algorithm can be summarized in the following steps \cite{Kaipio2006, calvetti2007, orlande2012inverse}:

\begin{algorithm}
\begin{algorithmic}[1]
\State Let $k = 0$ and start the Markov chain with sample $\textbf{P}^{(0)}$ at the initial state.
    \State Sample a candidate point $\textbf{P}^*$ from a proposal distribution $q(\textbf{P}^*|\textbf{P}^{(k)} )$.
    \State Compute direct problem from Equations (\ref{heateqDimless}) - (\ref{ICDimless}).
    \State Calculate the probability $\alpha(\textbf{P}^*|\textbf{P}^{(k)})$ with equation (\ref{alpha}).
    \State Generate a random value $U \thicksim \mathcal{U}(0,1)$, which is uniformly distributed in $(0,1)$.
    \State If $U \leq \alpha (\textbf{P}^*|\textbf{P}^{(k)})$, set $\textbf{P}^{k+1} = \textbf{P}^{(*)}$. Otherwise, set $\textbf{P}^{k+1} = \textbf{P}^{(k)}$.
    \State Make $k = k+1$ and return to step 2 in order to generate the sequence $\{\textbf{P}^{(1)} \,, \textbf{P}^{(2)} \,, ... \,, \textbf{P}^{(N_s)}\}$.
\end{algorithmic}
\end{algorithm}

\section{Case study}
\label{ssec:casestudy}

\subsection{Description}
\label{ssec:description}

\begin{figure}[!h]
\begin{center}
\subfigure[\label{fig:T_aL} air and a ground temperature at $x = L$]{\includegraphics[width=0.45\textwidth]{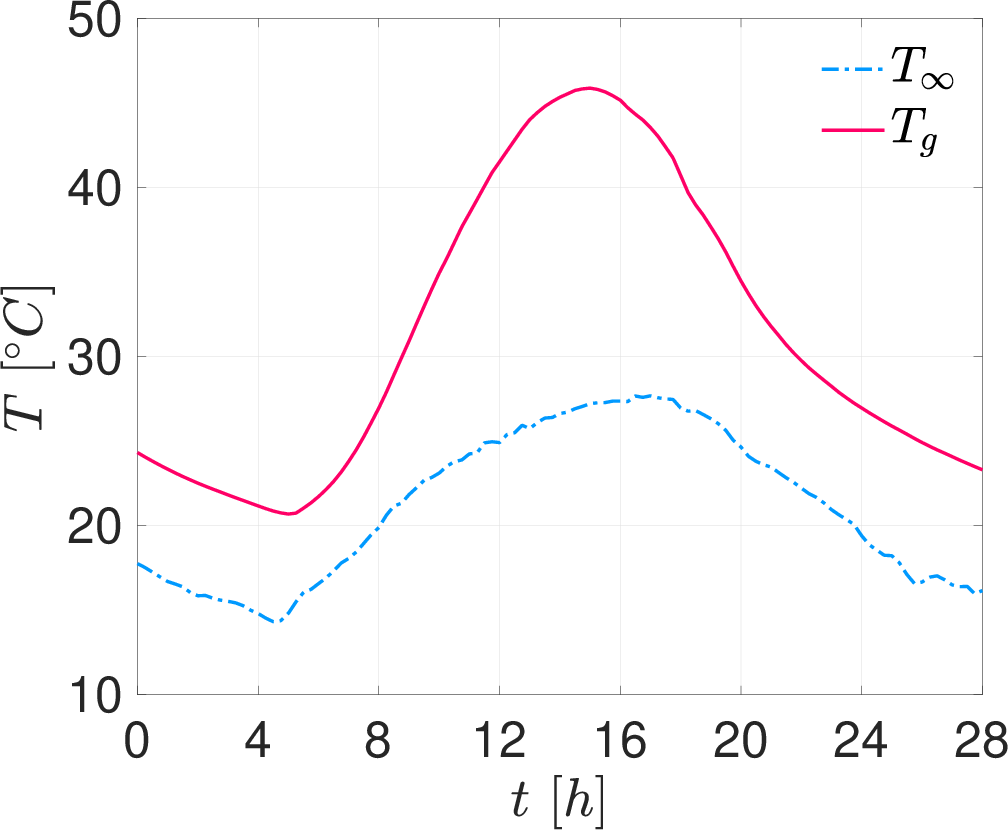}} \hspace{0.05cm}
\subfigure[\label{fig:Net} radiation heat flux]{\includegraphics[width=0.45\textwidth]{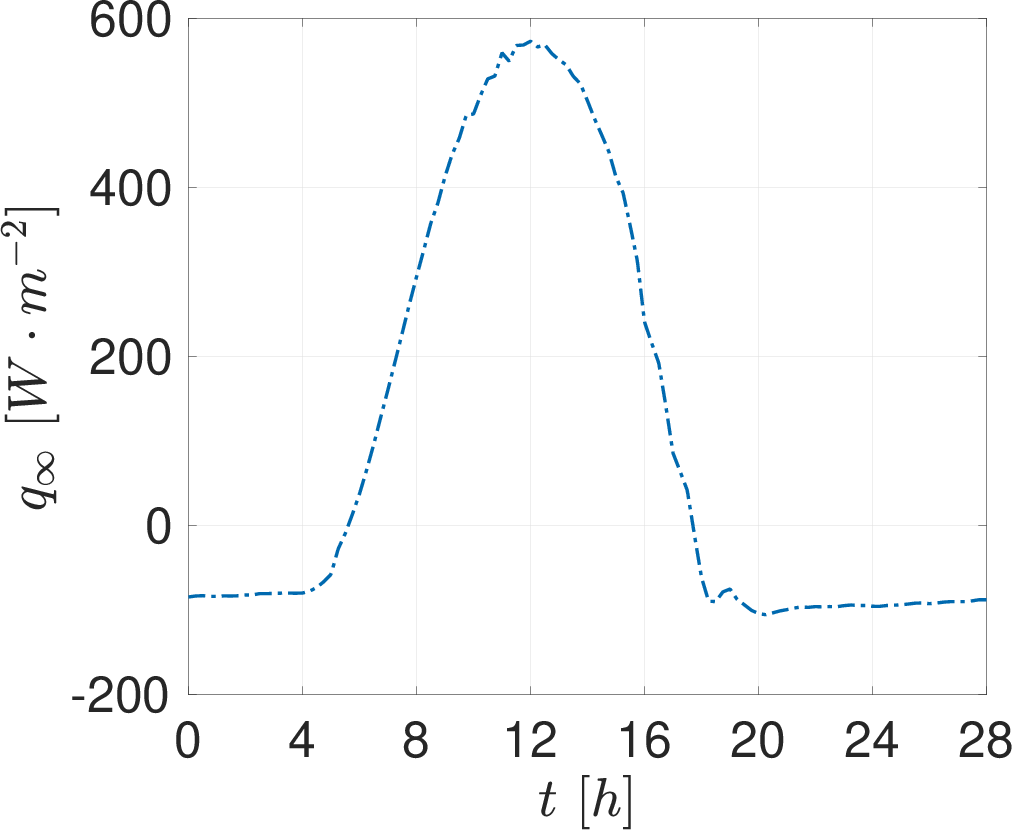}}\\
\subfigure[\label{fig:v} wind velocity and direction]{\includegraphics[width=0.5\textwidth]{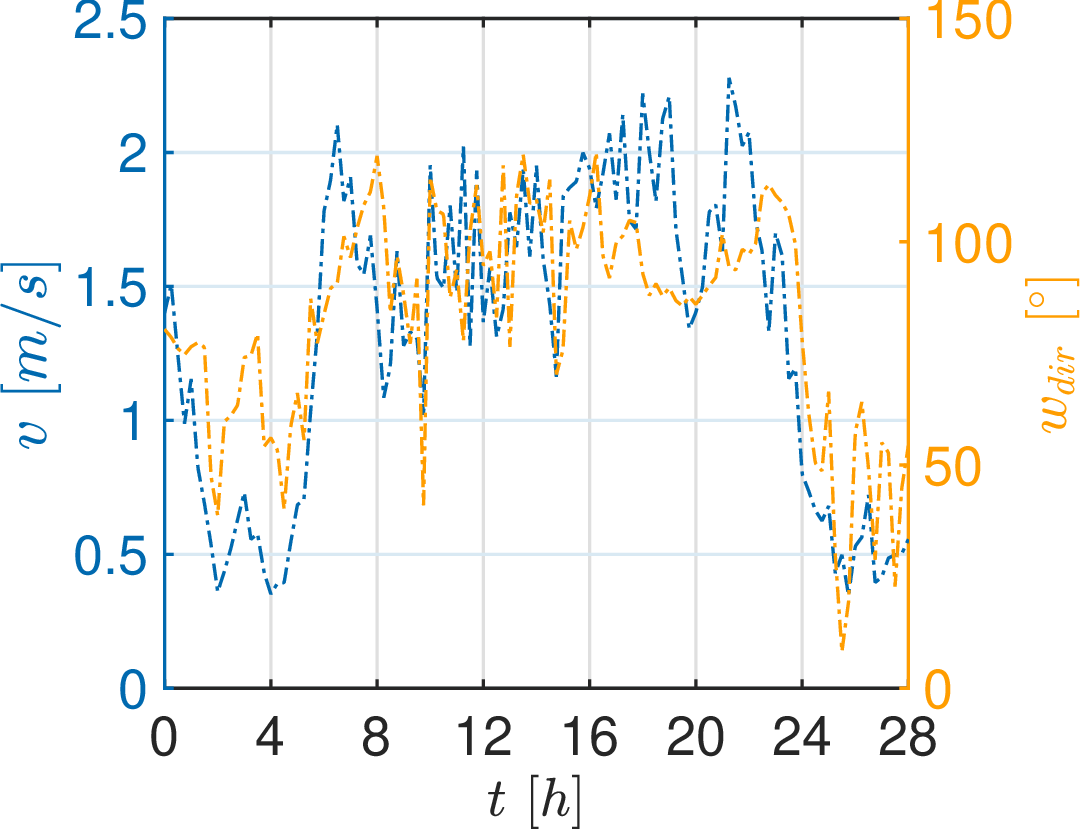}} \hspace{0.05cm}
\subfigure[\label{fig:T_in} initial temperature]{\includegraphics[width=0.45\textwidth]{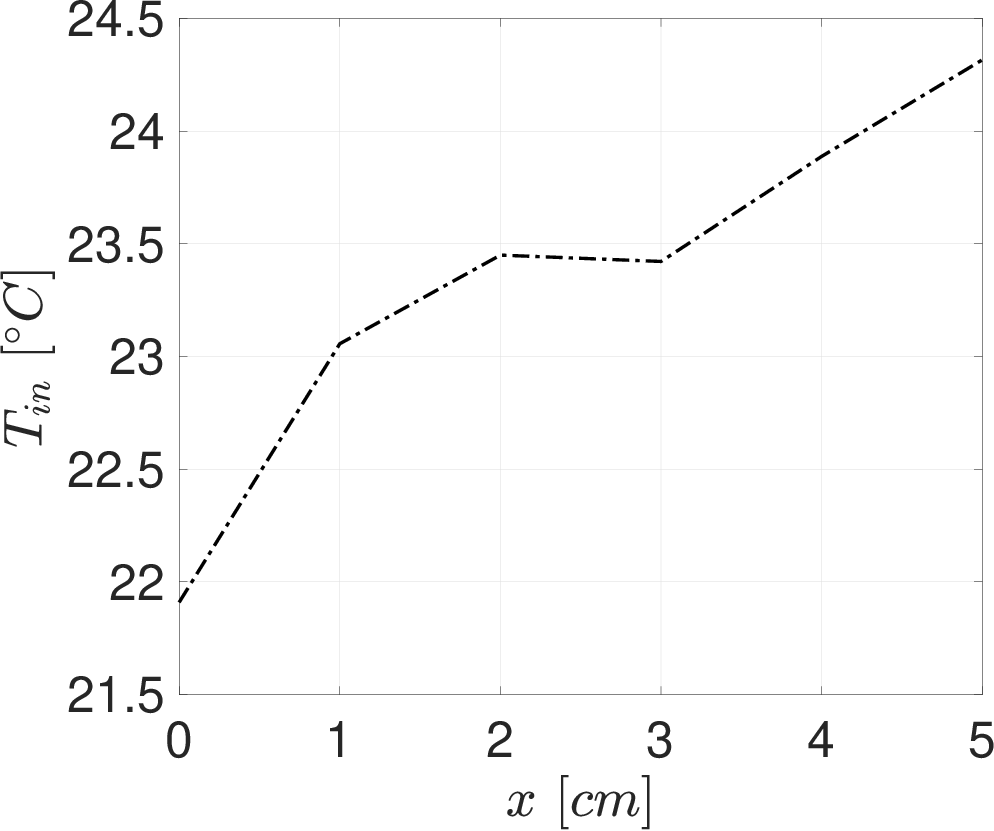}} \hspace{0.05cm}
 \caption{Time variation of the air and ground temperatures \emph{(a)}, radiation flux \emph{(b)}, wind velocity and direction \emph{(c)} and initial temperature \emph{(d)}.}
  \label{fig:BCs}
\end{center}
\end{figure}

The experimental data were provided by~\cite{COHARD2018675}, from experiments at a parking lot located within the Université Gustave Eiffel (former IFSTTAR Institute) in Bouguenais, France. The parking lot structure was composed of $5 \ \mathsf{cm}$ thick layer of asphalt concrete pavement, and a ballast layer under it. In the case study examined here, the total  depth is $L \egal 5 \ \mathsf{cm}$. Consequently, we should note that in the physical model we suppose the homogeneous thermal properties of the ground. Thermocouples were installed to measure the temperature at different ground depths. For measuring the surface temperature, $4$ thermocouples were used and infrared radiometer measurements were done in order to verify the impact of solar radiation and the accuracy of this measurement was considered as satisfying. The $M \egal 5$ sensors were positioned at $x_{\,m} \, \in \, \bigl\{\, 0\,, 1\,,\,2\,,\,3\,,\,4 \,\bigr\} \ \mathsf{cm}\,$. Measurements of the air and ground temperature at $x = L = 5 \ \mathsf{cm}$ are used for the boundary conditions~\eqref{BCL}--\eqref{BCR} of the direct problem. Their time variations given as $T_{\,\infty} (t)$ and $T_{\,g} (t)$ are shown in Figure~\ref{fig:T_aL}. 

In boundary condition \eqref{BCL} the $q_{\infty} (t)$ corresponds to the net radiation value from \cite{COHARD2018675}. The net radiation was measured by a net radiometer - NRlite (Kipp $\&$ Zonen). Its time variation is given in Figure \ref{fig:Net}. Moreover, Figure \ref{fig:v} presents the measurements of the wind velocity. As it can be seen, the wind velocity varies during the duration of the experiment, justifying that the physical model considers the surface heat transfer coefficient as a time dependent function. In the experiment, the radiation heat flux and the air temperature were measured at $1 \ \mathsf{m}$ above surface. The wind velocity was measured using a YOUNG sensor at $1.5 \ \mathsf{m}$ above the surface. Regarding the initial condition given in Figure \ref{fig:T_in}, it is obtained by interpolating the values of temperatures at each measurement positions at $t \egal 0\,$.

The initial experiments from \cite{COHARD2018675} were performed with a duration of $144 \ \mathsf{h}\,$ (on June 6 - 12, 2004) in the parking lot of  $2500 \ \mathsf{m^2}$ flat bare asphalt square. During the whole measurement period, only one natural rain was observed on June 10, and short drizzles on June 8 and 9. Since the phase change of moisture is not taken into account in the mathematical model, and no measurement of moisture content were carried out during the experimental campaign, the estimation procedure is carried out up to $t_{\,f} \egal 28 \ \mathsf{h}$ (just before the first rain event). 

\subsection{Practical identifiability}

\begin{figure}[!htb]
 \centering
\subfigure[\label{fig:SP_h1}
$h_1$]
{\includegraphics[width=.45\textwidth]{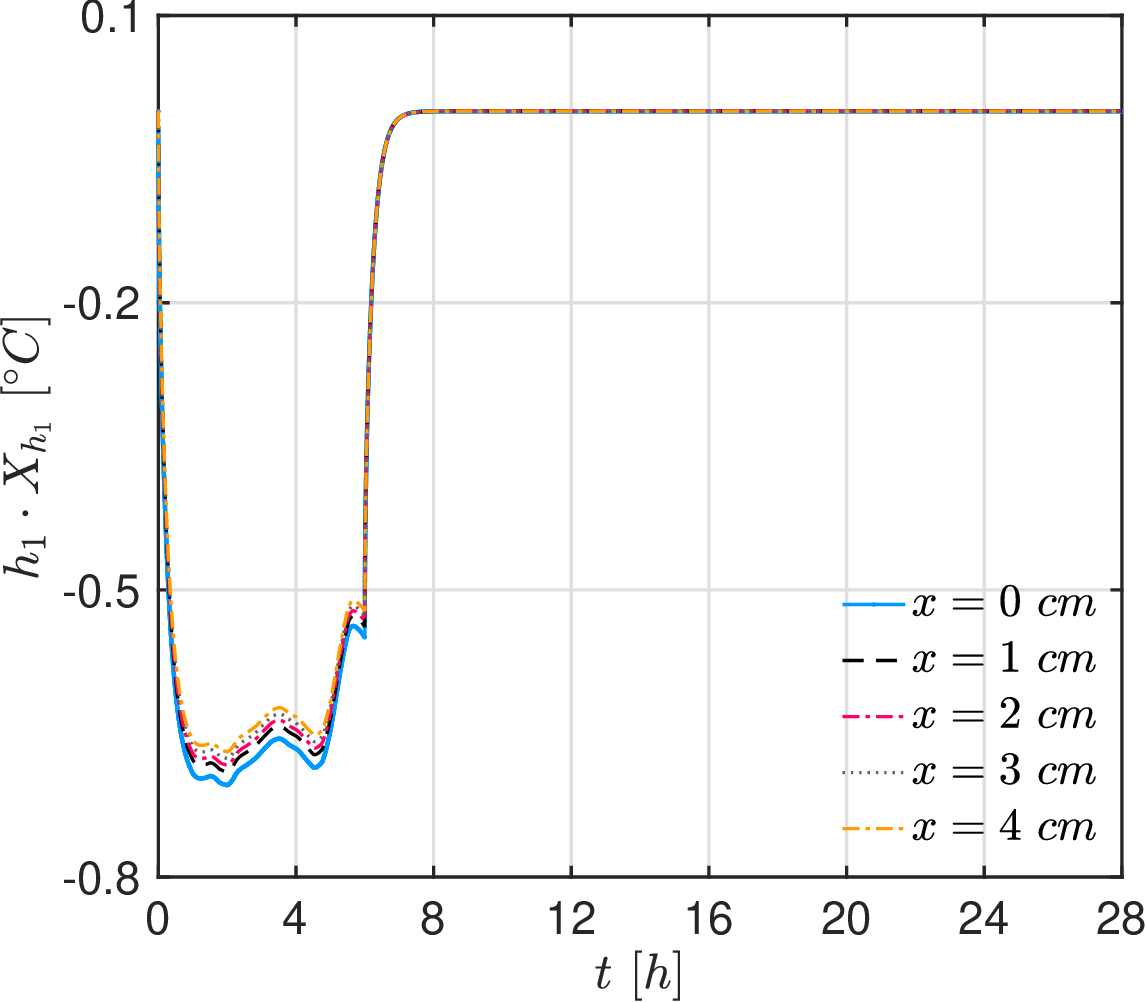}}  \hspace{0.05cm}
\subfigure[\label{fig:SP_h2}
$h_2$]
{\includegraphics[width=.45\textwidth]{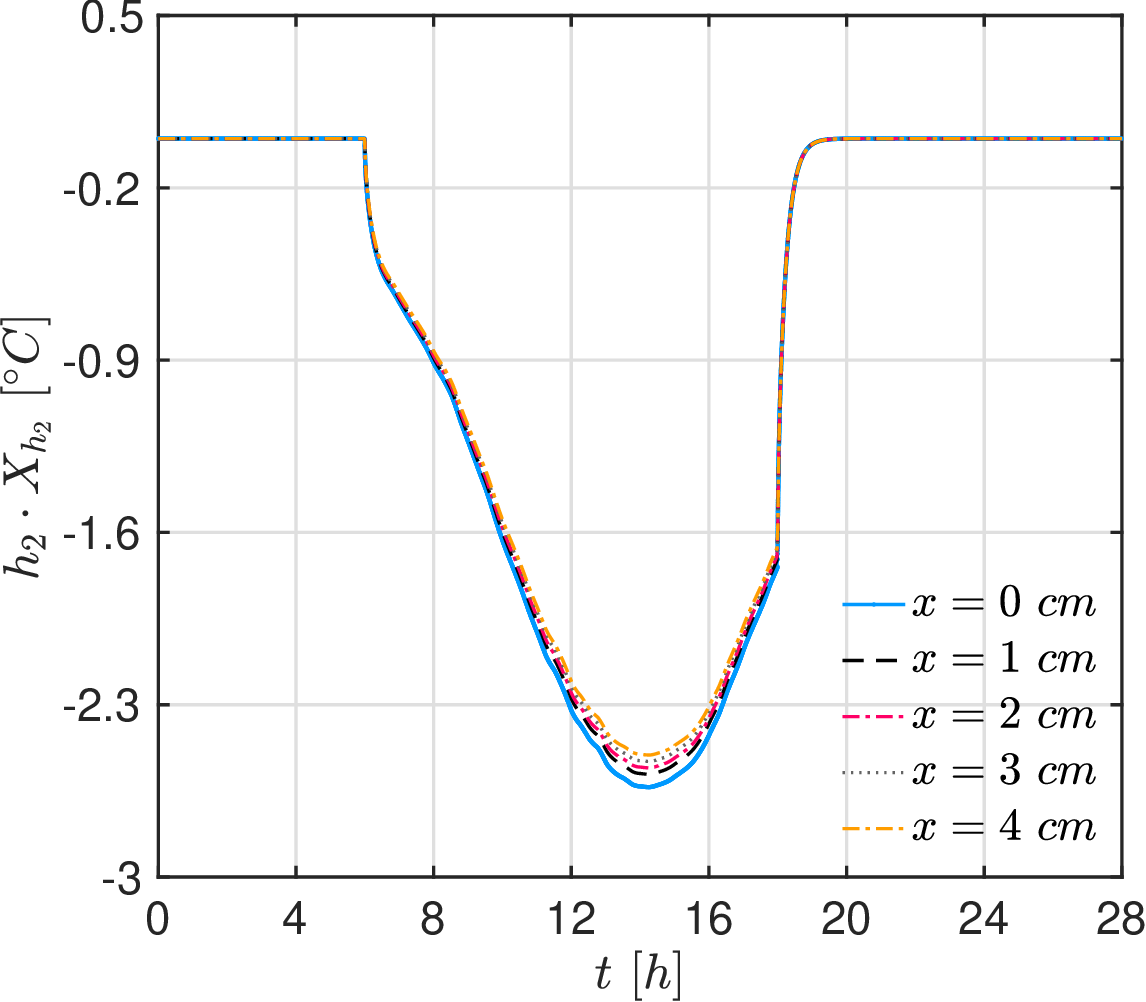}} \\
\subfigure[\label{fig:SP_h3}
$h_3$]{\includegraphics[width=.45\textwidth]{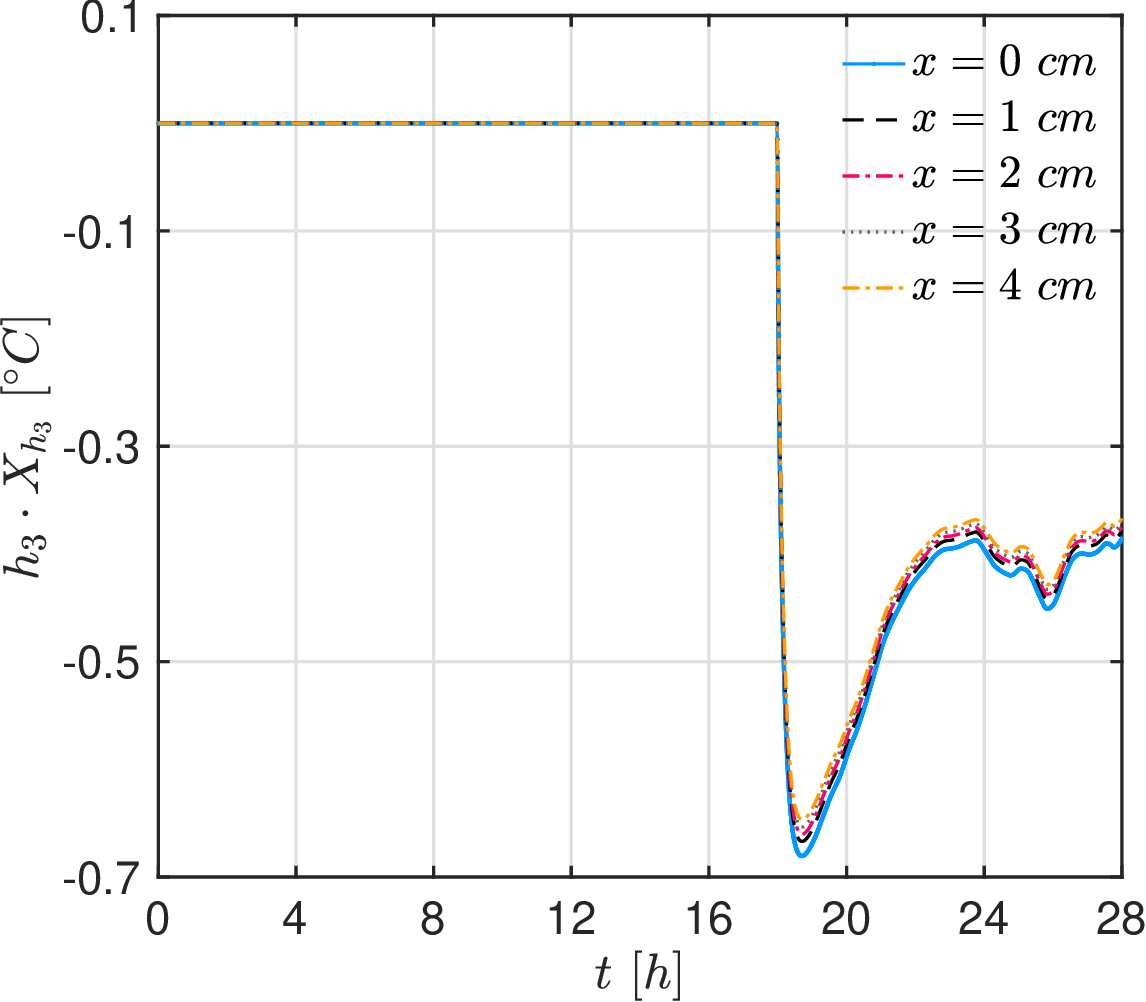}} 
\subfigure[\label{fig:SP_cv} $c_v$]
{\includegraphics[width=.45\textwidth]{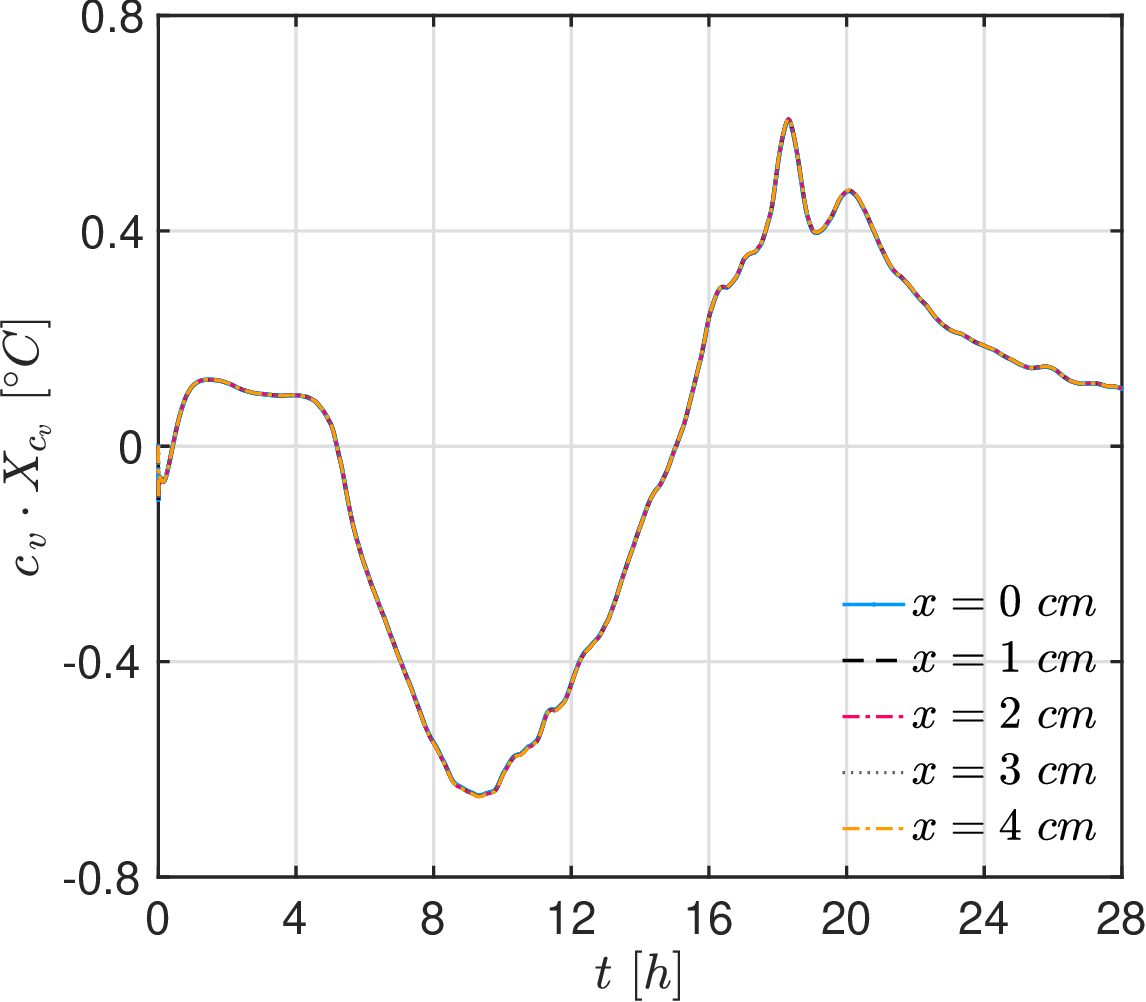}}  \\
\subfigure[\label{fig:SP_k} $\kappa$]
{\includegraphics[width=.45\textwidth]{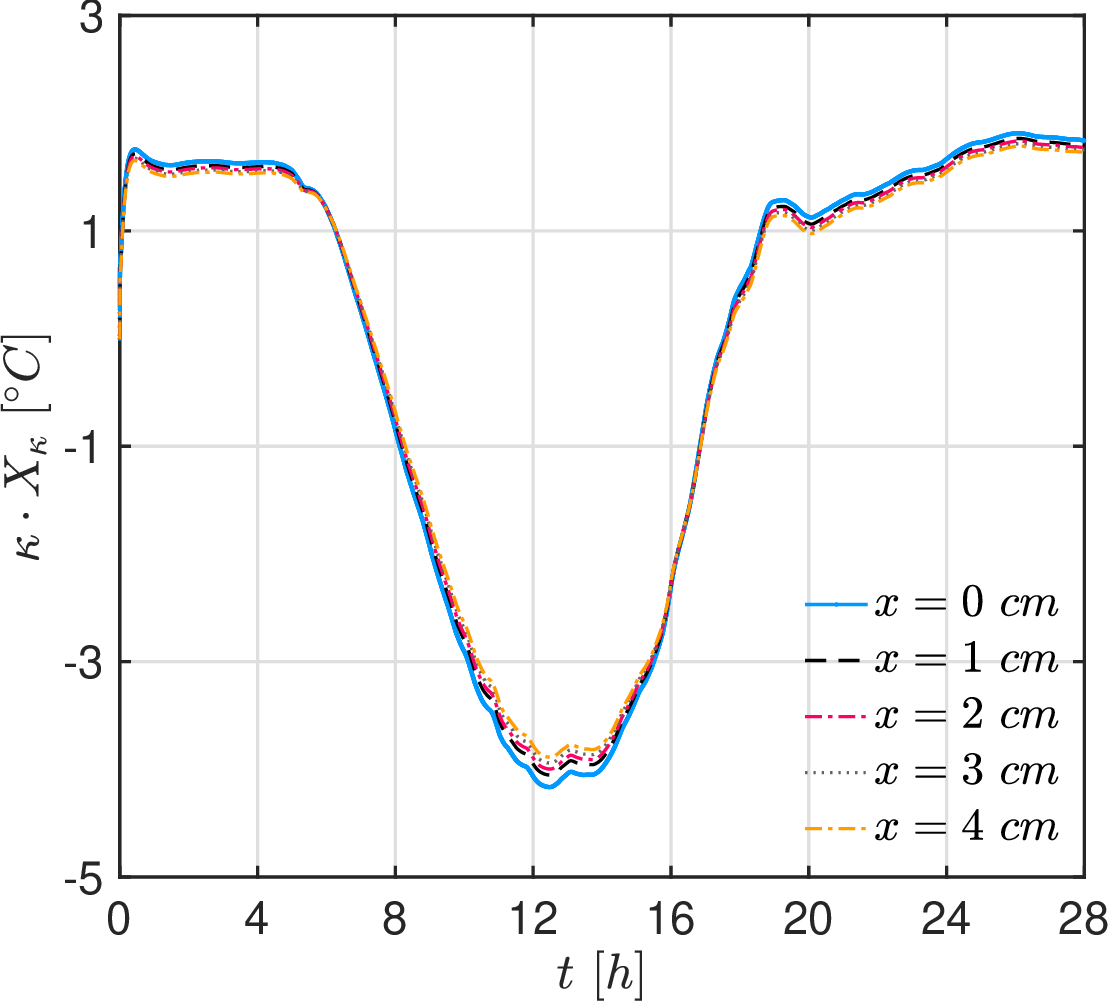}} 
\caption{Sensitivity coefficients for each parameter}.
\label{fig:SensP}
\end{figure}

The sensitivity coefficients are computed at each of the 5 sensor locations. From the literature review, the following values for the volumetric heat capacity and the thermal conductivity of the dry asphalt concrete are given $c_{\,v} = 2\,,1 \ \mathsf{MJ \cdot m^{-3}\cdot K^{-1}}$ and $\kappa = 2\,,23 \ \mathsf{W\cdot m^{-1} \cdot K^{-1}}$, respectively \cite{Incropera_2006}. 
Values of the surface heat transfer coefficients are taken as $h_{\,1} = h_{\,2} = h_{\,3} = 10 \  \mathsf{ W \cdot m^{-2} \cdot K^{-1}} $ for $Nt = 3$ within $t \in [0, 6] \, \mathsf{h}$, $t \in [6, 24] \, \mathsf{h}$, and $t \in [24, 28] \, \mathsf{h}$ time intervals  \cite{MIRSADEGHI2013134}.  Figures~\ref{fig:SP_h1} -~\ref{fig:SP_k} present the time variations of the reduced
sensitivity coefficients for $h_{\,1}$, $h_{\,2}$ and $h_{\,3}$, $C$ and $\kappa$, respectively. Note in Figures~\ref{fig:SP_h1} -~\ref{fig:SP_h3} that the sensitivity coefficients of $\left\{h_i\right\}_{i=1}^{i=3}$ are zero outside the corresponding time intervals where these parameters are not null. The sensitivity coefficients with respect to $h_{\,1}$ are large in magnitude, such as for $C$ and $\kappa$, and they are not linearly dependent. It is also interesting to note that the behaviors of the sensitivity coefficients were not affected by the sensor position, probably due to the small depth of ground considered. 

\subsection{Results of the parameter estimation problem}

The Metropolis-Hastings algorithm is applied for solving the inverse problem. Results of the three case studies are presented in the next subsections.

In Case A, the surface heat transfer coefficients, $h_1$, $h_2$ and $h_3$, are defined for the time intervals between $t \in [0, 6] \, \mathsf{h}$, $t \in [6, 18] \, \mathsf{h}$, and $t \in [18, 28] \, \mathsf{h}$, respectively. These time intervals are chosen based on analysis of Figure \ref{fig:Net}, when the net radiation flux have different magnitudes. In Case B, the time periods of $t \in [0, 6] \, \mathsf{h}$, $t \in [6, 24] \, \mathsf{h}$, and $t \in [24, 28] \, \mathsf{h}$ are chosen, based on the wind velocity presented in Figure \ref{fig:v} which shows different levels during these time intervals.

In Cases A and B, independent Gaussian priors with a positive constraint are used for all parameters. Mean and standard deviations of the thermal properties and surface heat transfer coefficients for the Gaussian priors presented in Table \ref{tab:Initial guess}.

In Case C, the surface heat transfer coefficient was discretized as piecewise constant values within each time interval when the measurements were taken (every $15 \ \mathsf{min}$). Thus, for Case C a Gaussian Markov Random Field prior was used for the parameter values representing the time variation of the surface heat transfer coefficient. The same Gaussian priors of Case A were specified for the volumetric heat capacity and thermal conductivity. The hyperprior for the parameter $\gamma$ in the Gaussian Markov Random Field prior was modeled as a Rayleigh distribution given by Equation \ref{Rayleigh_distr}, with $\gamma_0$ set as $2 \,.22 \,  \, \mathsf{m^{4} \cdot K^{2} \, \cdot W^{-2}}$.

\subsubsection{Cases A and B}

In this section results of the estimation problem for both Cases A and B are presented.

\begin{figure}[!htb]
\begin{center}
\subfigure[\label{fig:cv_ab}]{\includegraphics[width=0.45\textwidth]{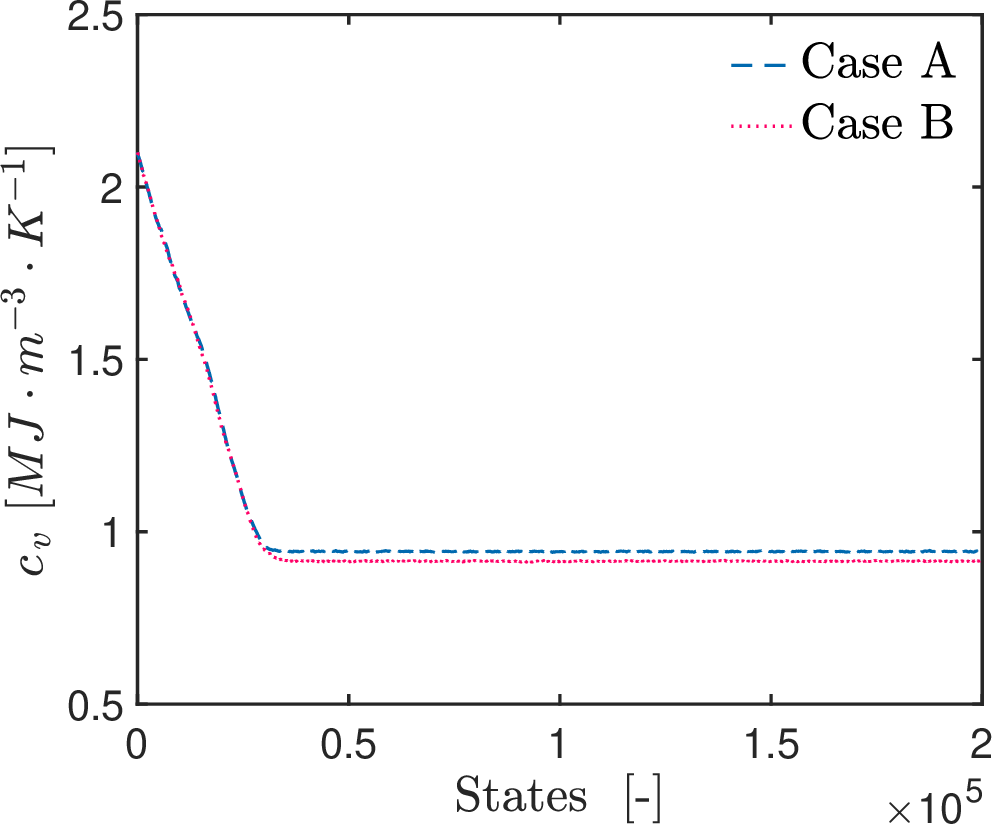}} \hspace{0.2cm}
\subfigure[\label{fig:cv_hist_ab}]{\includegraphics[width=0.45\textwidth]{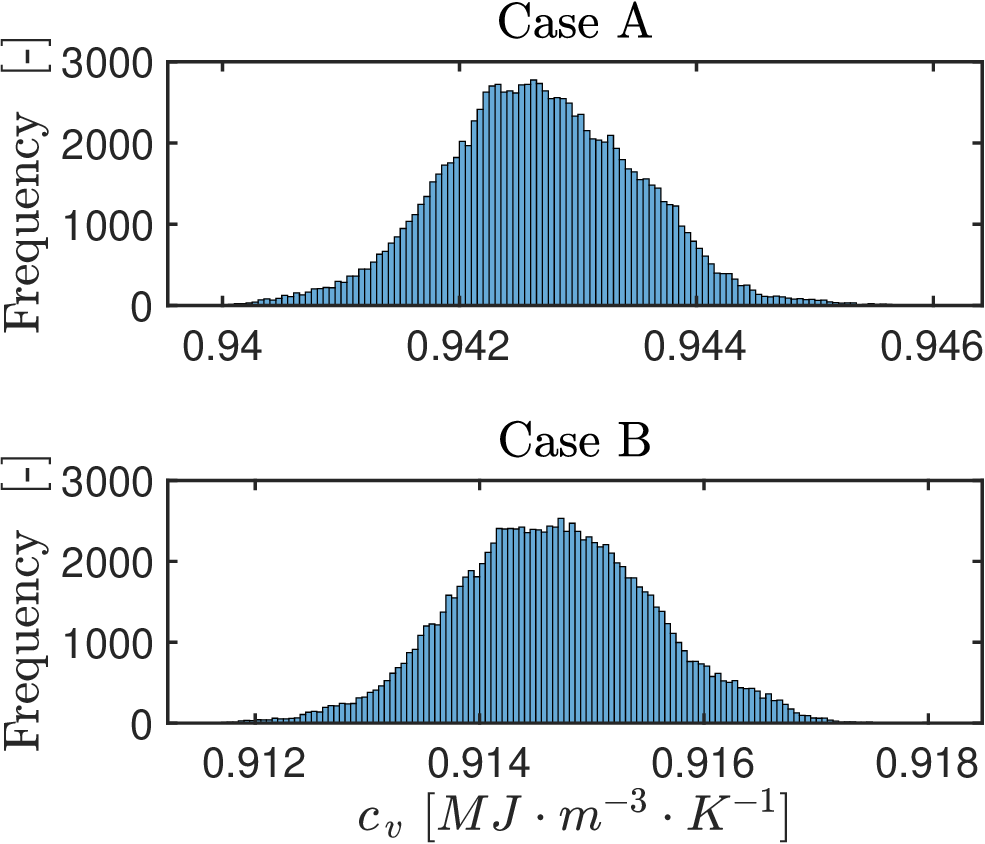}} 
\\
\subfigure[\label{fig:k_ab}]{\includegraphics[width=0.45\textwidth]{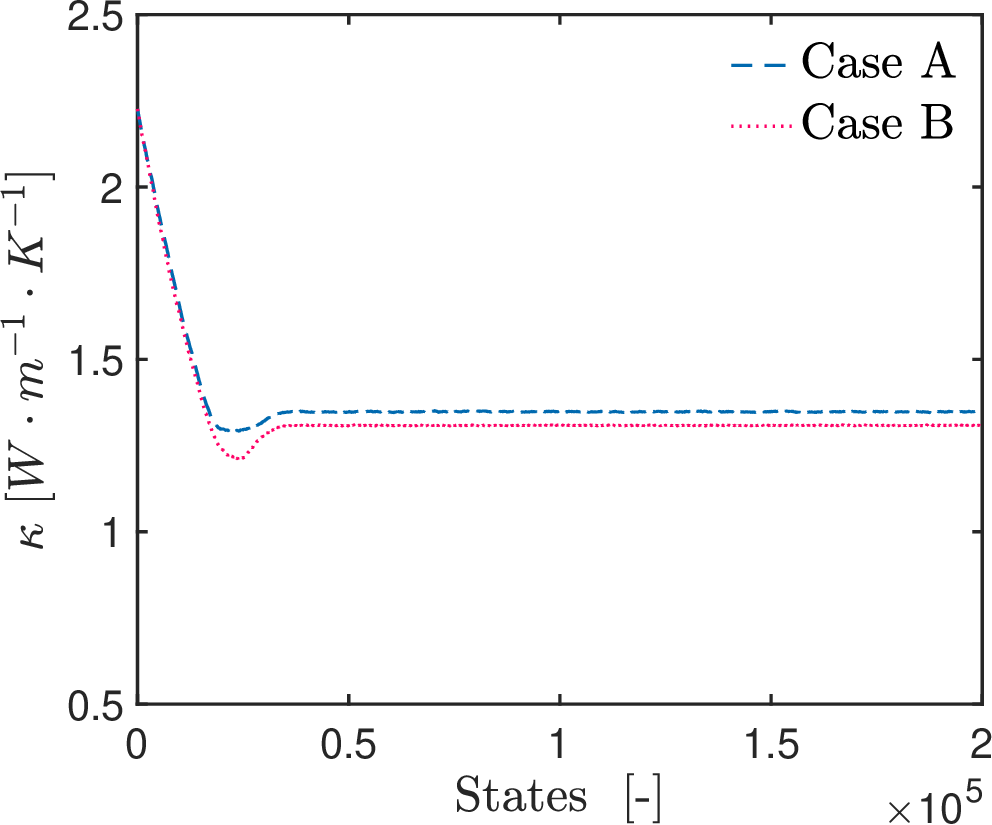}} \hspace{0.2cm}
\subfigure[\label{fig:k_hist_ab}]{\includegraphics[width=0.45\textwidth]{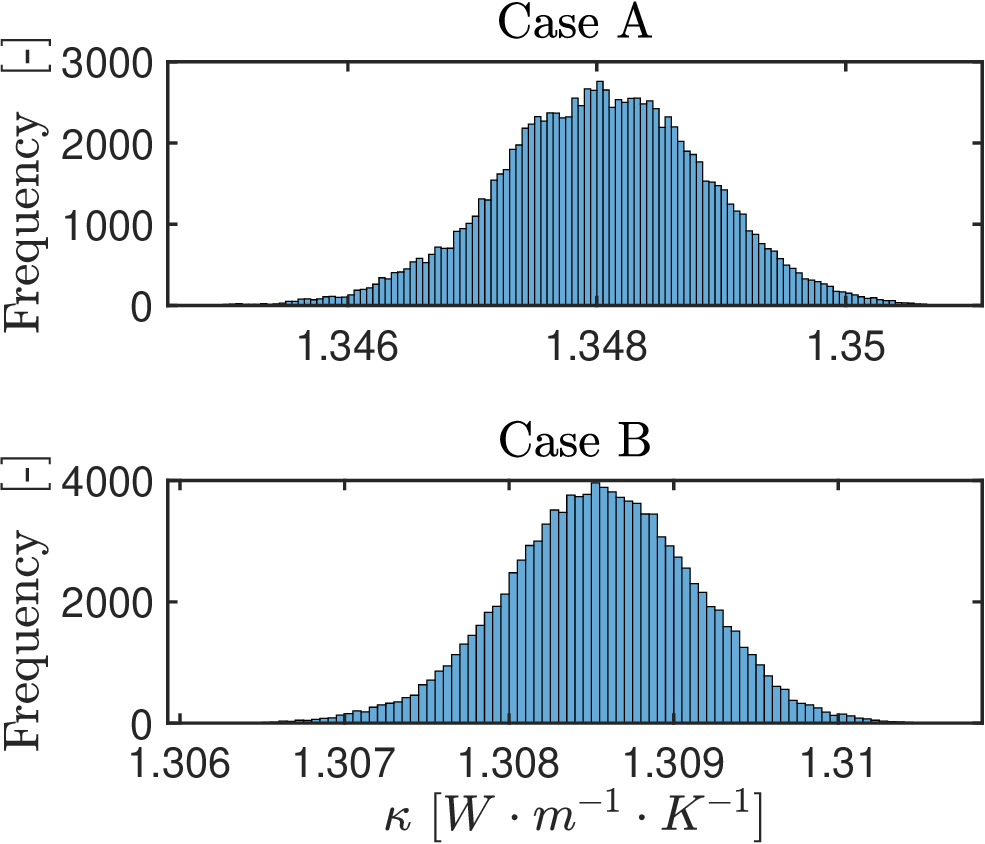}} 
\caption{Markov chains for: $c$ \emph{(a)} and $\kappa$ \emph{(c)}, Histograms of the samples after the burn-in for: $c$ \emph{(b)} and $\kappa$ \emph{(d)} for both cases A and B}
\label{fig:IP_parameters_ab}
\end{center}
\end{figure}

\begin{figure}[!htb]
\begin{center}
\subfigure[\label{fig:ht_ab}]{\includegraphics[width=0.45\textwidth]{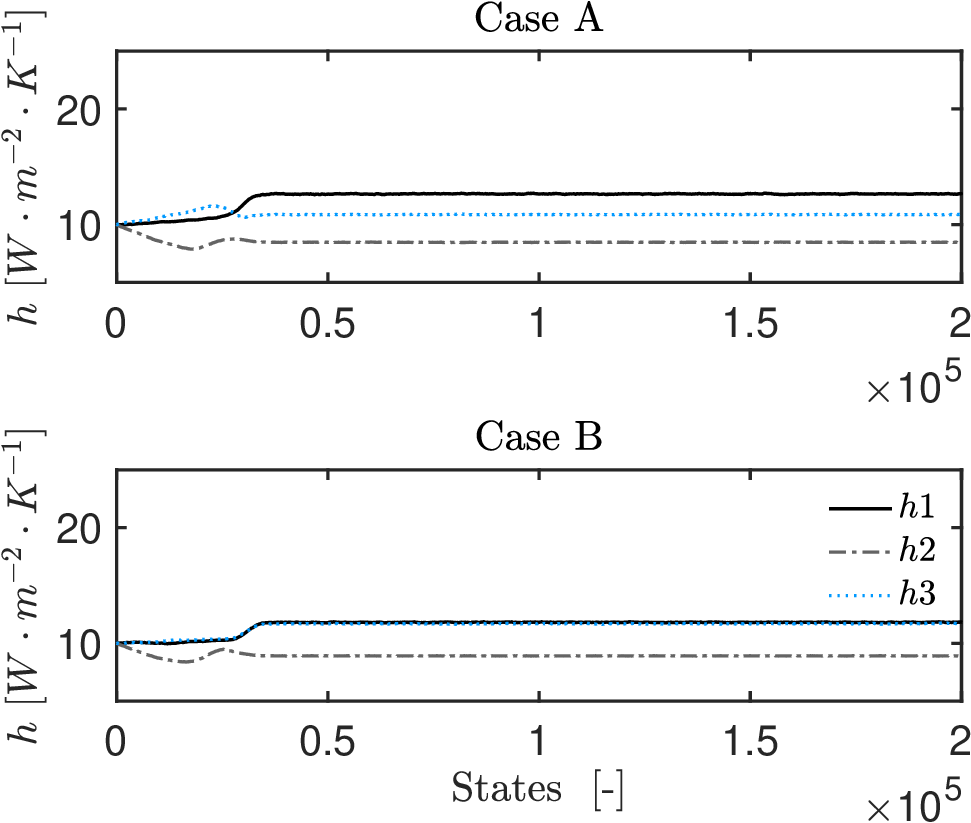}} \hspace{0.2cm}
\subfigure[\label{fig:h1_hist_ab}]{\includegraphics[width=0.45\textwidth]{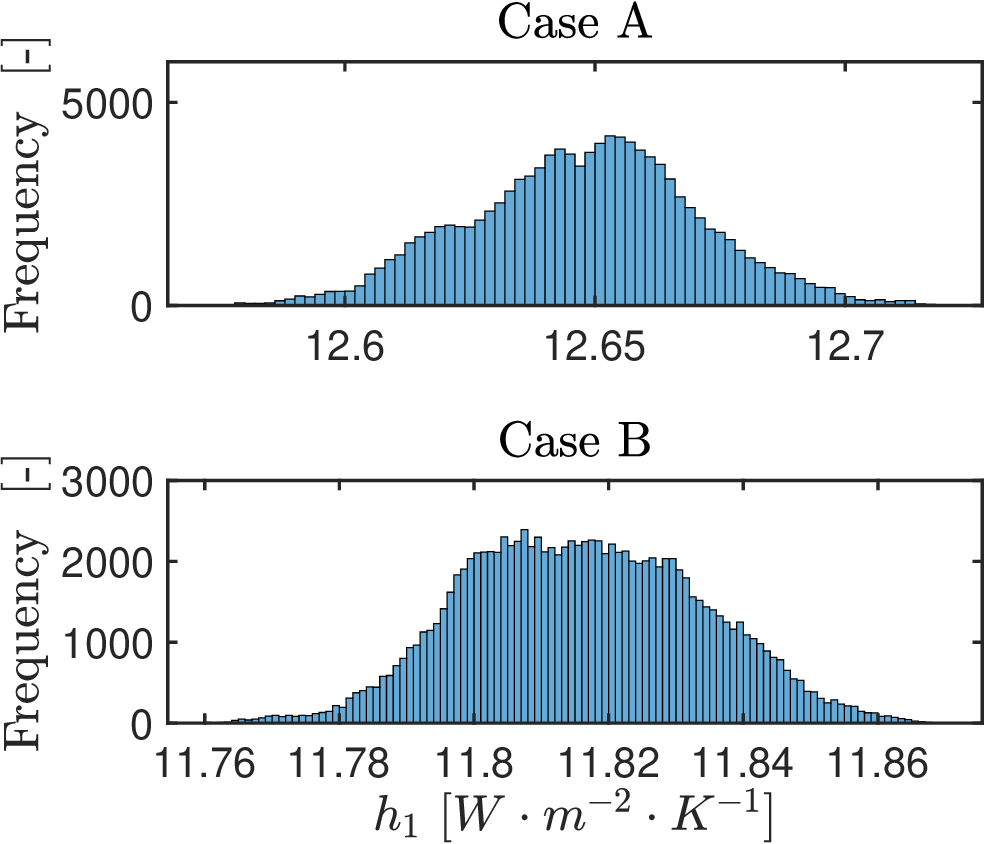}} 
\\
\subfigure[\label{fig:h2_hist_ab}]{\includegraphics[width=0.45\textwidth]{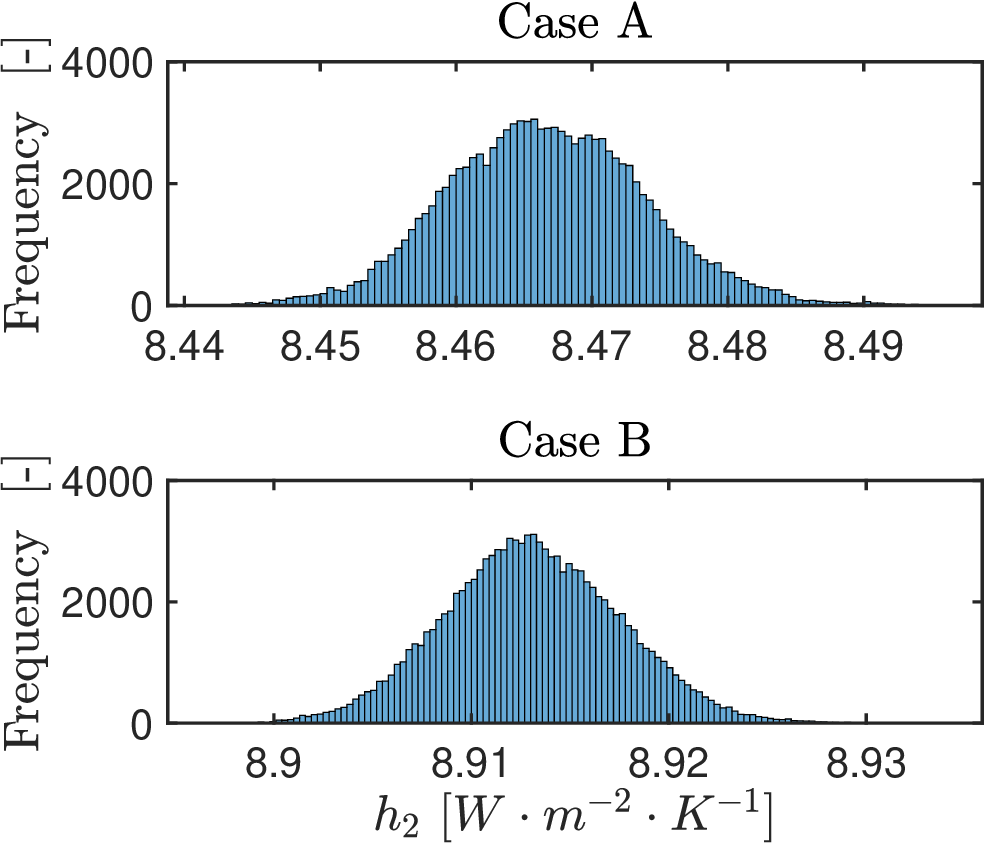}} \hspace{0.2cm}
\subfigure[\label{fig:h3_hist_ab}]{\includegraphics[width=0.45\textwidth]{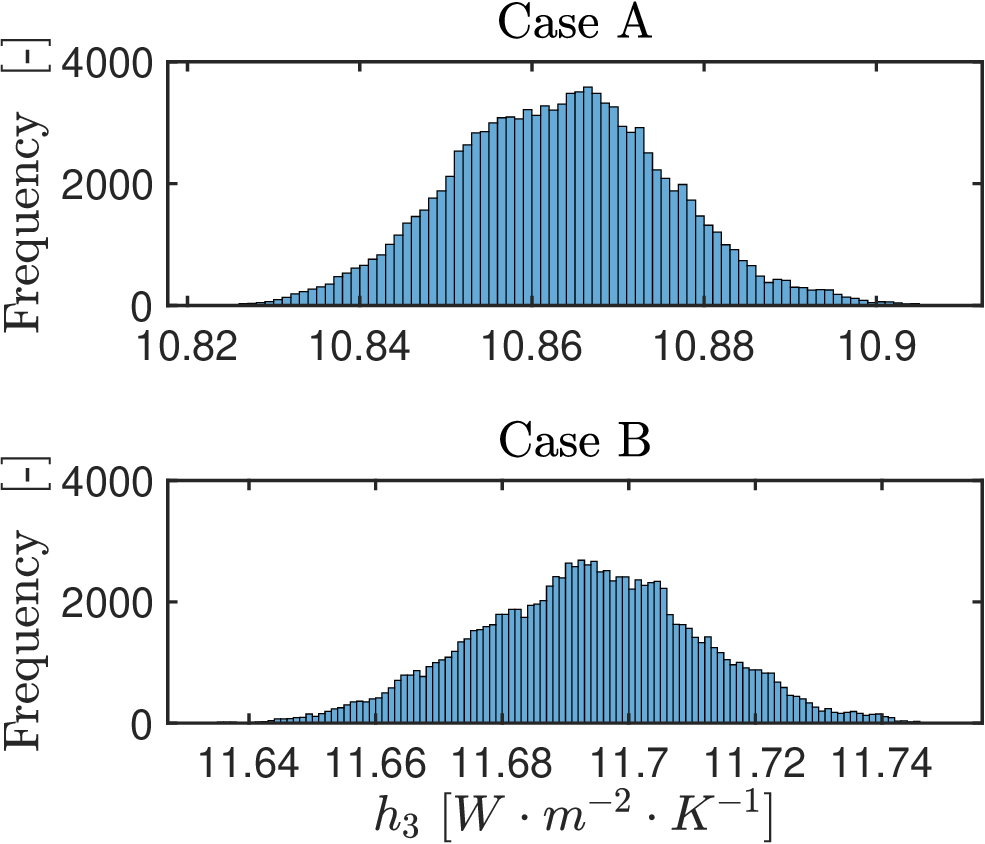}} 
\\
\caption{ Markov chains for: $\left\{h_i\right\}_{i=1}^{i=3}$ \emph{(a)}. Histograms of the samples after the burn-in for: $h_1$ \emph{(b)},  $h_2$ \emph{(c)}, and $h_3$ \emph{(d)} for both cases A and B.}
\label{fig:h_par_a}
\end{center}
\end{figure}

Figures \ref{fig:cv_ab} - \ref{fig:k_hist_ab} present the Markov chains and the histograms after the burn-in period for $C$ and $\kappa$, in Cases A and B. Similar results are presented in Figures \ref{fig:ht_ab} - \ref{fig:h3_hist_ab} for $h_1$, $h_2$ and $h_3$. It shows that the Markov chains, simulated with 200 000 states, reached equilibrium distributions after about 50 000 states, so that the burn-in period is taken as 90 000 states.

The histograms of the samples in the Markov chains after the burn-in period are shown in Figures \ref{fig:cv_hist_ab} and \ref{fig:k_hist_ab} for the volumetric heat capacity and the thermal conductivity, respectively, and Figures \ref{fig:h1_hist_ab}, \ref{fig:h2_hist_ab}, and \ref{fig:h3_hist_ab} for the surface heat transfer coefficients. These histograms present Gaussian distributions. Also the mean value of each unknown variable calculated after the burn-in period is computed. Those mean values will be used later for direct problem computations with estimates for residual investigations. Table \ref{tab:Initial guess} presents the statistics of the priors and of the marginal posteriors for Cases A and B. Literature values commonly used for the parameters \cite{Incropera_2006} are also shown in Table \ref{tab:Initial guess}. The estimated mean values for the parameters for Cases A and B have similar values with a relative error between $5$  to $7\%$ 
for $h_1$, $h_2$, $h_3$, and around $3\%$
for $c_v$, $\kappa$, respectively. 

\begin{figure}[!htb]
\begin{center}
\subfigure[\label{fig:acceptans_ab}]{\includegraphics[width=0.4\textwidth]{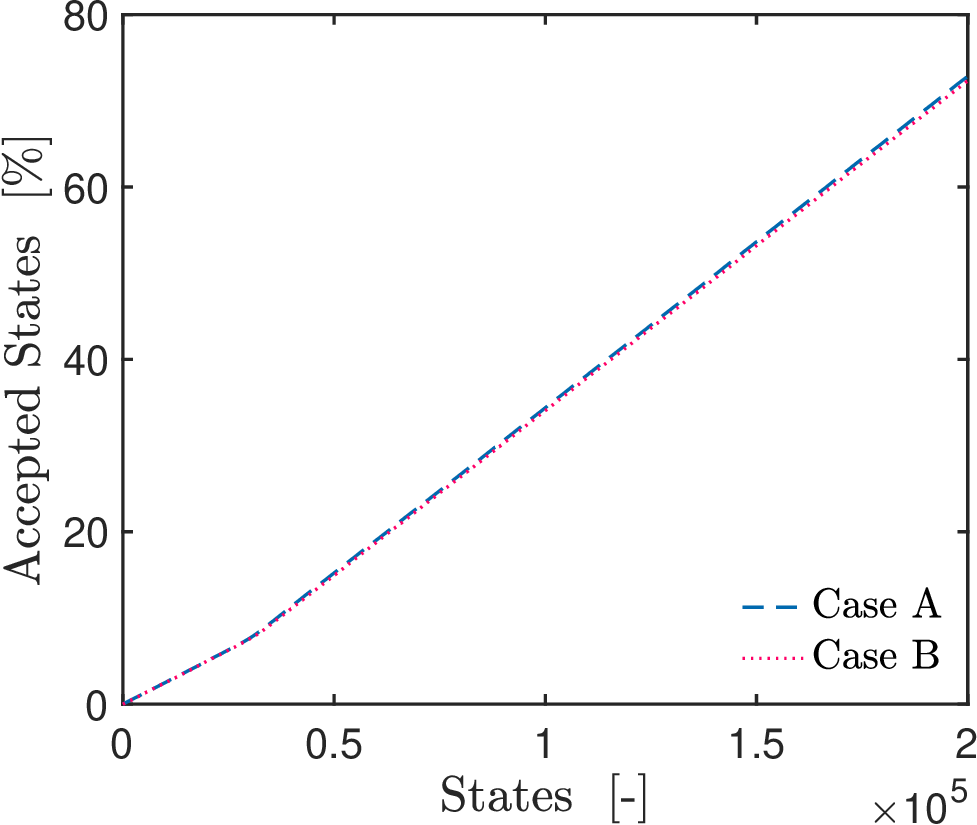}} \hspace{0.2cm}
\subfigure[\label{fig:likelihood_ab}]{\includegraphics[width=0.4\textwidth]{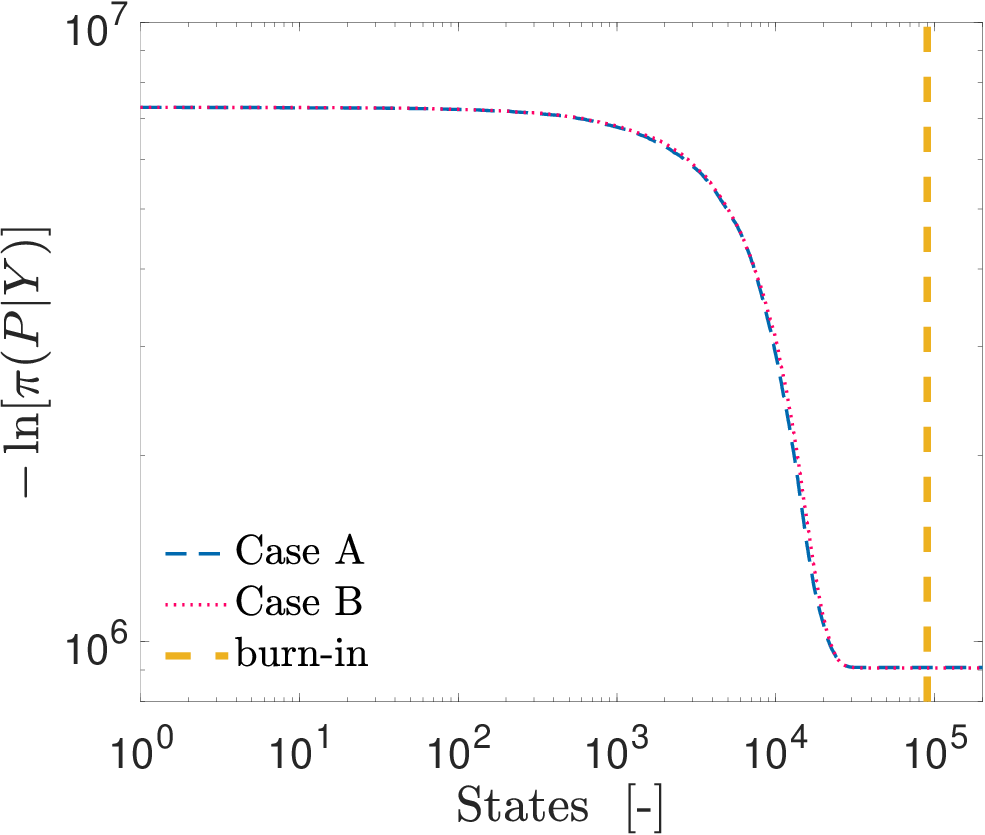}}  
\caption{Number of candidates accepted \emph{(a)}, Variation of the posterior distribution \emph{(b)}}
\label{fig:accepts_post}
\end{center}
\end{figure}

\begin{table}[!htb]
\centering
\caption{Statistics of the priors and of the marginal posteriors of the parameters for Cases A and B.}
\label{tab:Initial guess}
\setlength{\extrarowheight}{.5em}
\begin{tabular}{|c|c|c|c|c|c|c|}
\hline
\hline 
& & $h_1$ & $h_2$ & $h_3$ & $c_{\,v}$ & $\kappa$ \\[5pt]
& & $\unit{W\cdot m^{-2} \cdot K^{-1}}$ & $ \unit{W\cdot m^{-2} \cdot K^{-1}}$ & $\unit{W\cdot m^{-2} \cdot K^{-1}}$ & $\unit{MJ \cdot m^{-3}\cdot K^{-1}}$ & $\unit{W\cdot m^{-1} \cdot K^{-1}}$ \\[5pt]
\hline
lit. val. & -- & $10$ & $10$ & $10$ & $2 \,.10$ & $2 \,.27$\\
\hline
Prior & mean  & $10$ & $10$ & $10$ & $2 \,.10$ & $2 \,.27$\\
& std  & $5$ & $5$ & $5$ & $0 \,.021$ & $0 \,.1135$\\
\hline
 Case A & mean & $12 \,.65$ & $8 \,.47$ & $10 \,.86$ &  $0 \,. 94$ & $1 \,.35$ \\
 (burn-in $9 \cdot 10^4$)& std & $0 \,.02$ & $0 \,.01$ & $0 \,.01$ &  $0 \,. 0008$ & $0 \,. 0008$  \\
\hline
 Case B & mean & $11 \,.82$ & $8 \,.91$ & $11 \,.696$ &  $0 \,. 91$ & $1 \,.31$ \\
(burn-in $9 \cdot 10^4$) & std & $0 \,.02$ & $0 \,.0044$ & $0 \,.02$ &  $0 \,. 0009$ & $0 \,. 0006$  \\
\hline
\hline
\end{tabular}
\end{table}

The number of candidates generated with the proposal distribution that are accepted in step 5 of the Metropolis-Hastings algorithm is presented in Figure \ref{fig:acceptans_ab}. The acceptance rate is about $70 \%$ for cases A an B.
Figure \ref{fig:likelihood_ab} shows that the maximum a posteriori objective function decreases with increasing number of states of the Markov chain. It can be seen that after the burn-in period, which is indicated by the yellow vertical line,  the Markov chain of the posterior distribution converged to an equilibrium distribution.

\begin{figure}[!htb]
\begin{center} 
\vspace*{-2.0cm}
\subfigure[\label{fig:h_apr_est_a} $h(\,t\,)$]{\includegraphics[width=0.45\textwidth]{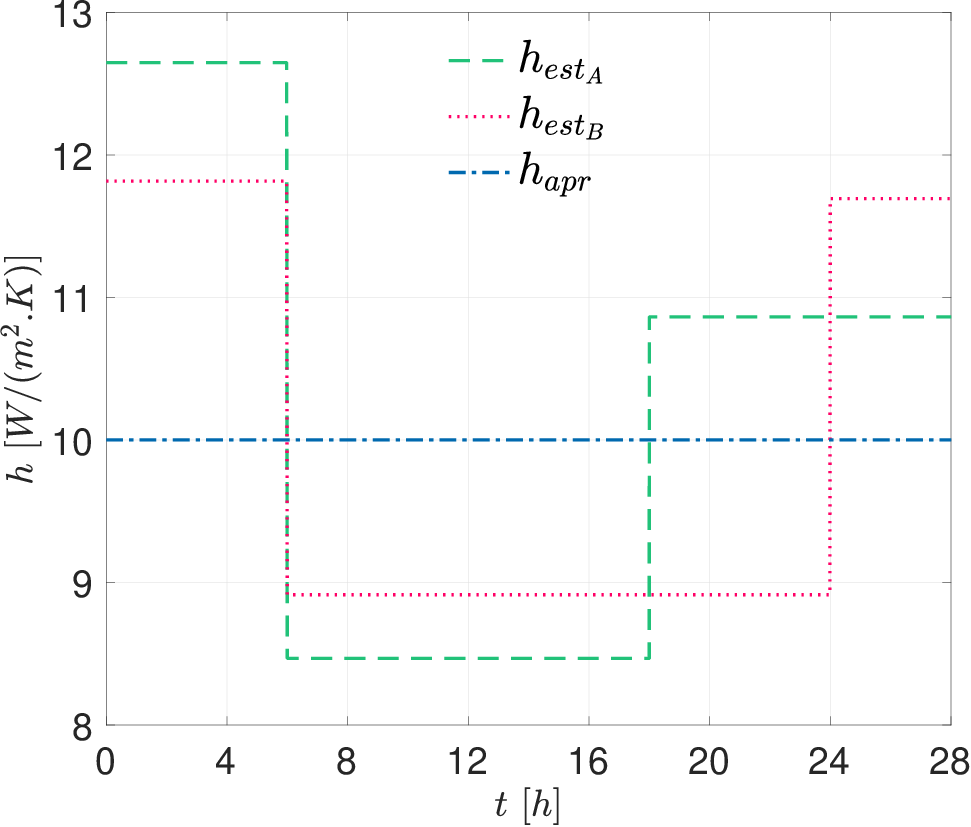}} \hspace{0.2cm}
\subfigure[\label{fig:Tx0_a} $x_m = 0 \ \mathsf{cm}$]{\includegraphics[width=0.45\textwidth]{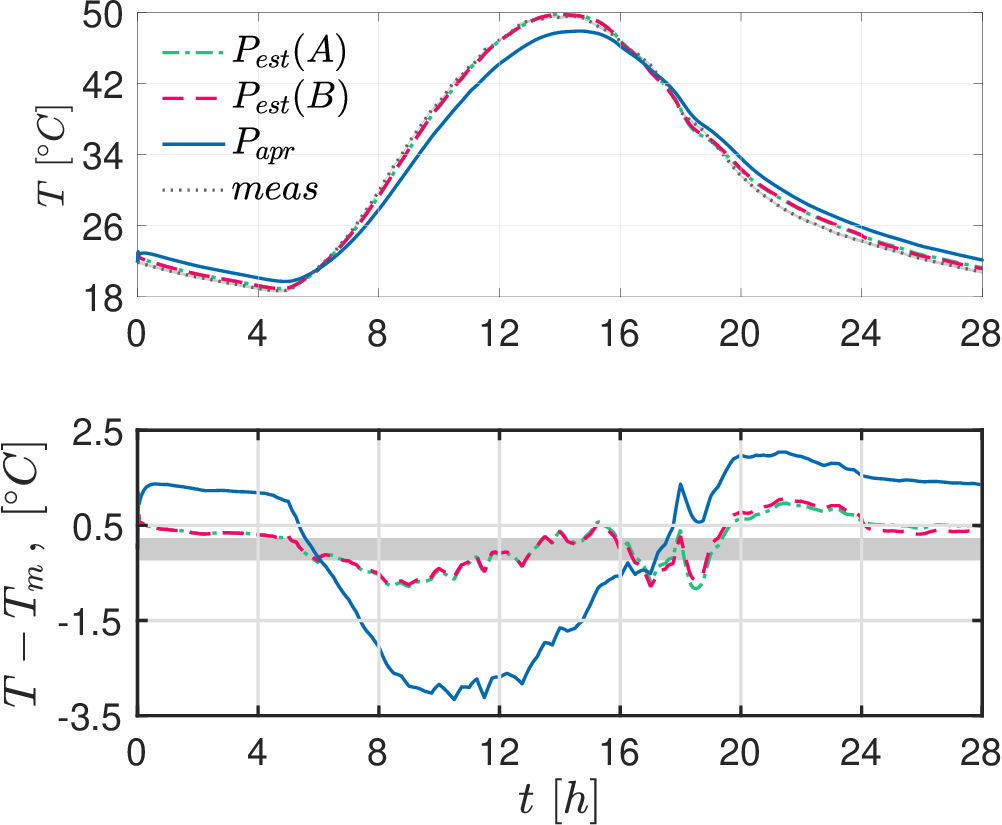}} \\
\subfigure[\label{fig:Tx1_a} $x_m = 1 \ \mathsf{cm}$]{\includegraphics[width=0.45\textwidth]{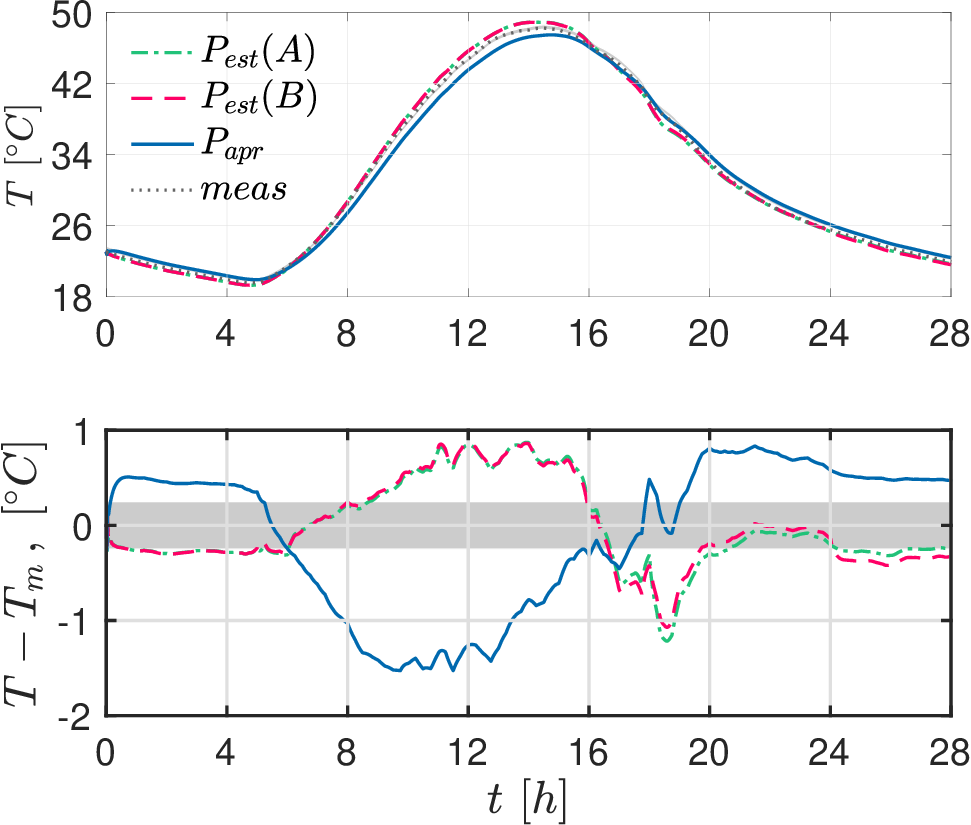}} \hspace{0.2cm}
\subfigure[\label{fig:Tx2_a} $x_m = 2 \ \mathsf{cm}$]{\includegraphics[width=0.45\textwidth]{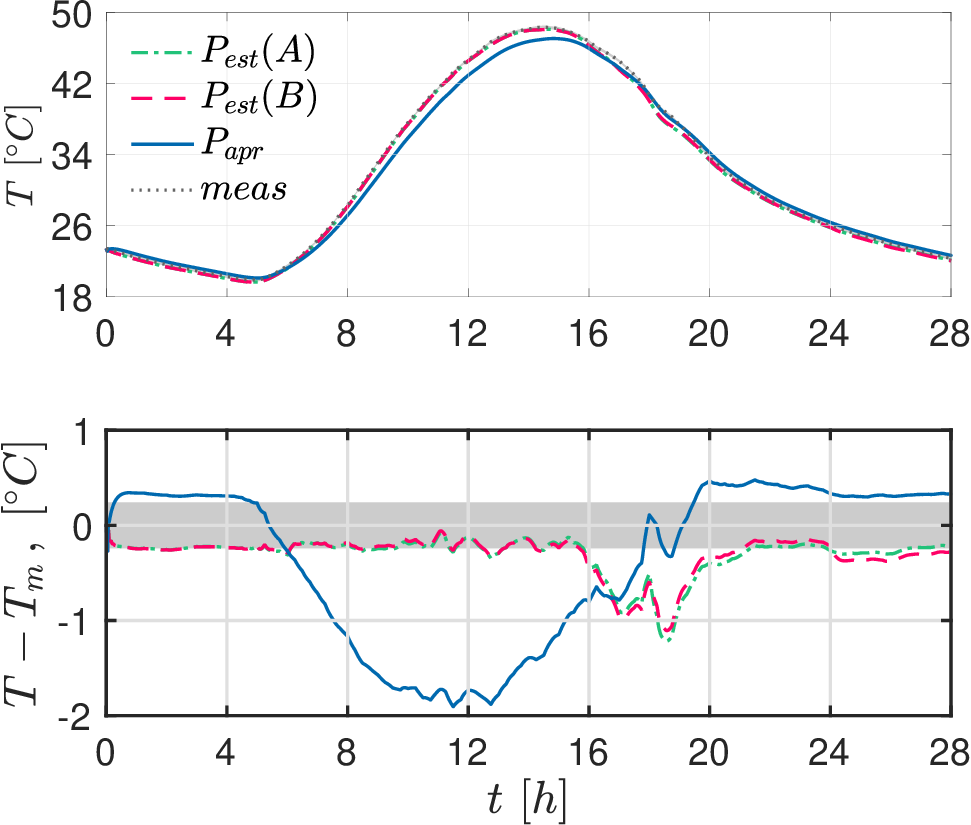}} \\
\subfigure[\label{fig:Tx3_a} $x_m = 3 \ \mathsf{cm}$]{\includegraphics[width=0.45\textwidth]{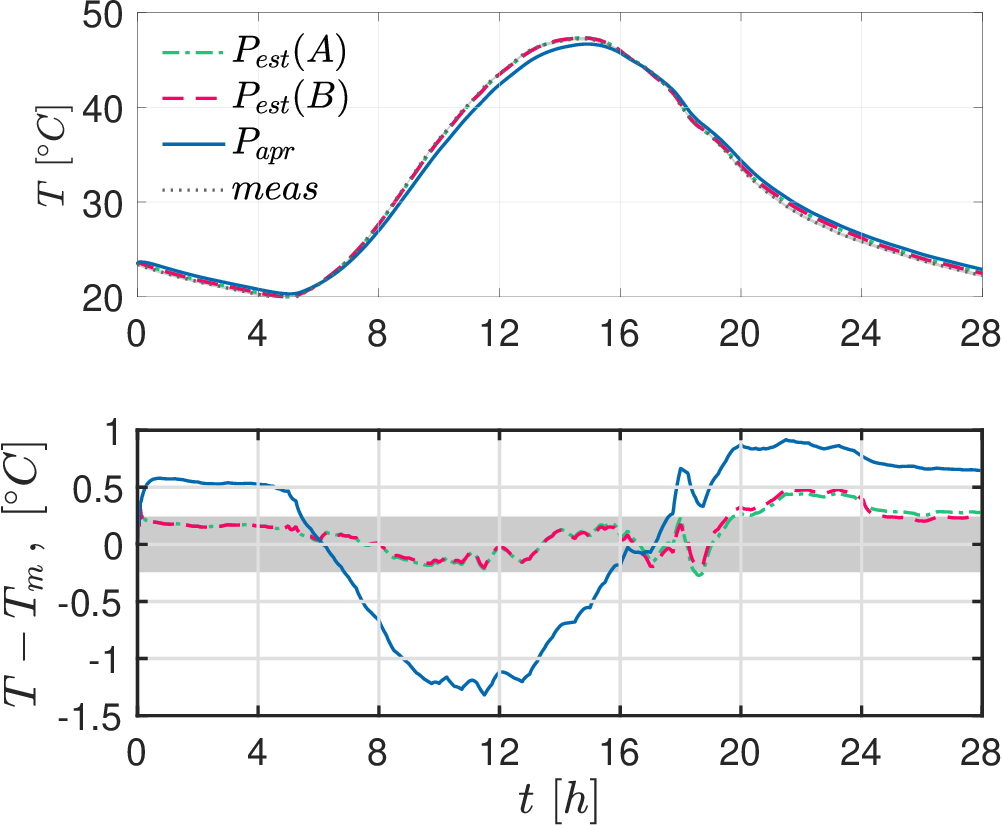}} \hspace{0.2cm}
\subfigure[\label{fig:Tx4_a} $x_m = 4 \ \mathsf{cm}$]{\includegraphics[width=0.45\textwidth]{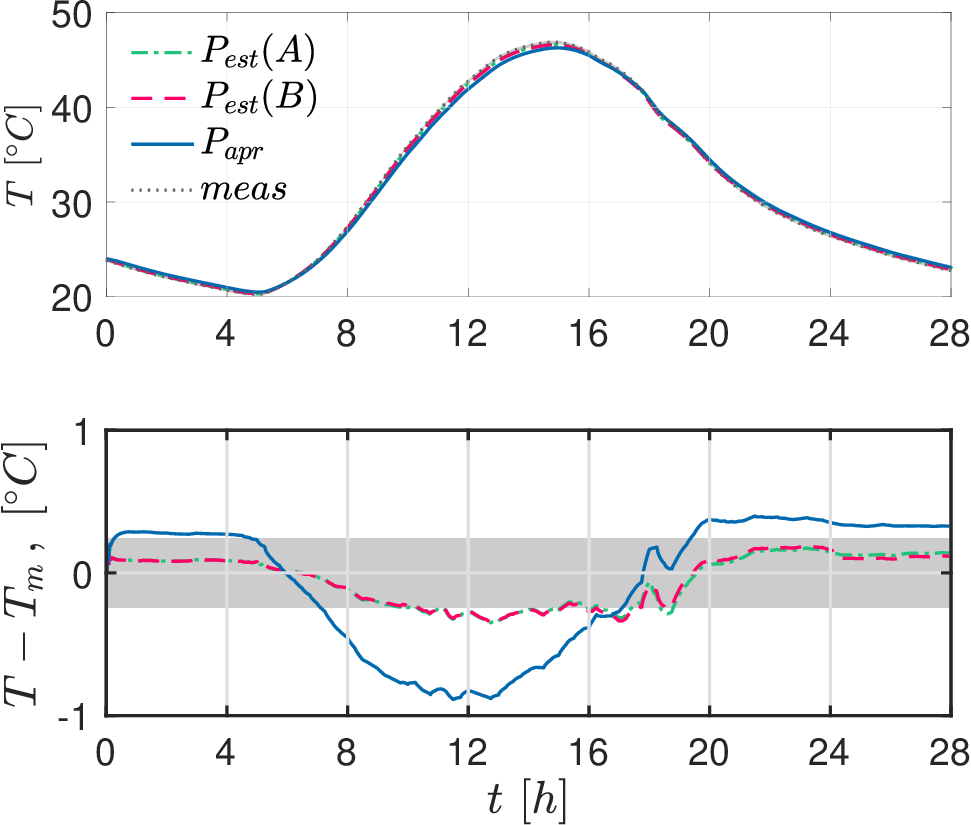}} \\
\caption{ Time varying surface heat transfer coefficient \emph{(a)}. Computed temperatures with mean of the estimations and literature values, and measured values of temperature at several sensor locations \emph{(b)} - \emph{(f)}. }
\label{fig:DP_results_a}
\end{center}
\end{figure}

Figure \ref{fig:h_apr_est_a} presents the mean of posterior distributions and \textit{a priori} time dependent surface heat transfer coefficient obtained with the inverse problem solution for Cases A and B. As we mentioned earlier, in case A the total time is divided into three periods, according to the net radiation flux variations. As it can be seen from Figure \ref{fig:h_apr_est_a} and Figure \ref{fig:Net}, the lowest value of the net radiation flux variation corresponds to the largest value of the surface heat transfer coefficient. In case B, the total time was also divided into to three periods, according to the wind velocity behavior. Figure \ref{fig:h_apr_est_a} and Figure \ref{fig:v} show that the surface heat transfer coefficient is inversely proportional to the wind velocity. This is contrary to \emph{empirical} law proposed in the literature \cite{MIRSADEGHI2013134}. Such behavior is probably due to the fact that wind and temperature measurements are carried at different heights in the experimental design. Furthermore, given the magnitude of the temperature difference between the ground surface and the ambient air during the day, natural convection effects may be more important than forced one. In general, Figure \ref{fig:h_apr_est_a} shows that cases A and B have similar behavior of the surface heat transfer coefficient on the given time intervals.

After the estimation procedure, the direct problem is solved with the mean values of the marginal posterior distributions  of thermal conductivity, volumetric heat capacity, and surface heat transfer coefficients, as well as with the literature values presented in Table \ref{tab:Initial guess}. The corresponding temperatures computed at five measurement positions and the respective measurements are compared in Figures~\ref{fig:Tx0_a} - \ref{fig:Tx4_a}. In the figures showing the residuals, defined by the differences between measurements and predicted temperatures computed with estimated parameters, the grey areas show the standard deviation of the measurements, given by $0.25 \, \mathsf{^\circ C}$. For both cases, residuals obtained with the mean parameters estimated with the solution of the inverse problems were lower than the residuals obtained with the literature values with a maximum relative discrepancies of $0.07 \%$, $0.93 \%$, $2.44 \%$, $6.94 \%$ and $8.83 \%$ at $x = 0 \, \mathsf{cm}$, $x = 1 \, \mathsf{cm}$, $x = 2 \, \mathsf{cm}$, $x = 3 \, \mathsf{cm}$ and $x = 4 \, \mathsf{cm}$, respectively. Note that the residuals are correlated and larger than the standard deviation of the measurements, thus indicating that the mathematical model with the parameters estimated in Cases A and B still demand further improvement for the representation of the experimental data. The time variation of the surface heat transfer coefficient is examined in the section below. While the surface heat transfer coefficient was discretized with three piecewise constant values in Cases A and B, a much more refined discretization is used for Case C, with piecewise constant values during each time interval where the measurements were available, that is, every $15 \ \mathsf{min}$.

\subsubsection{Case C} 

In this section results of the inverse problem, obtained considering a Gaussian prior for $\kappa$ and $c_v$, and a Gaussian smoothness prior for the time varying surface coefficient $h(t)$ are presented. As a reminder, $N_{\,t}\egal 113$ for the modeling of the surface coefficient Eq.~\eqref{h(t)}.

\begin{figure}[!htb]
\begin{center}
\subfigure[\label{fig:res_c}]{\includegraphics[width=0.45\textwidth]{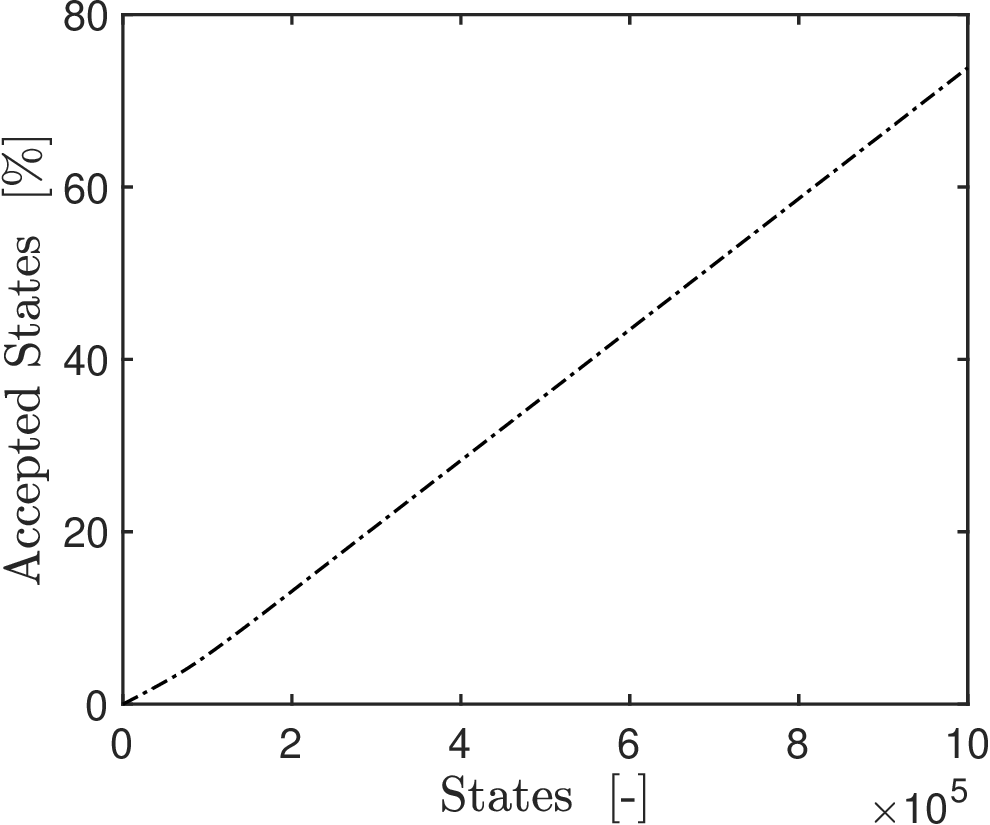}} \hspace{0.2 cm}
\subfigure[\label{fig:posterior_c}]{\includegraphics[width=0.45\textwidth]{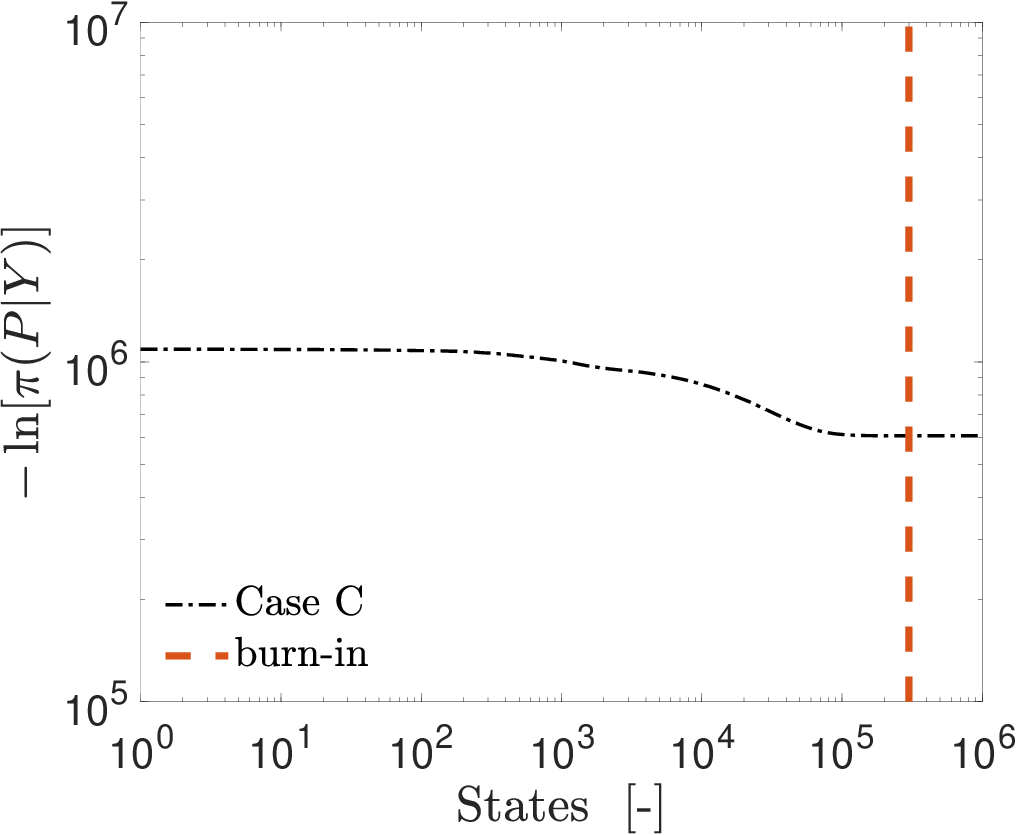}}
\caption{Number of accepted  candidates \emph{(a)} and variation of the posterior distribution \emph{(b)} according to the number of states.}
\label{fig:accep_s_poster_c}
\end{center}
\end{figure}

\begin{figure}[!htb]
\begin{center}
\subfigure[\label{fig:cv_c}]{\includegraphics[width=0.45\textwidth]{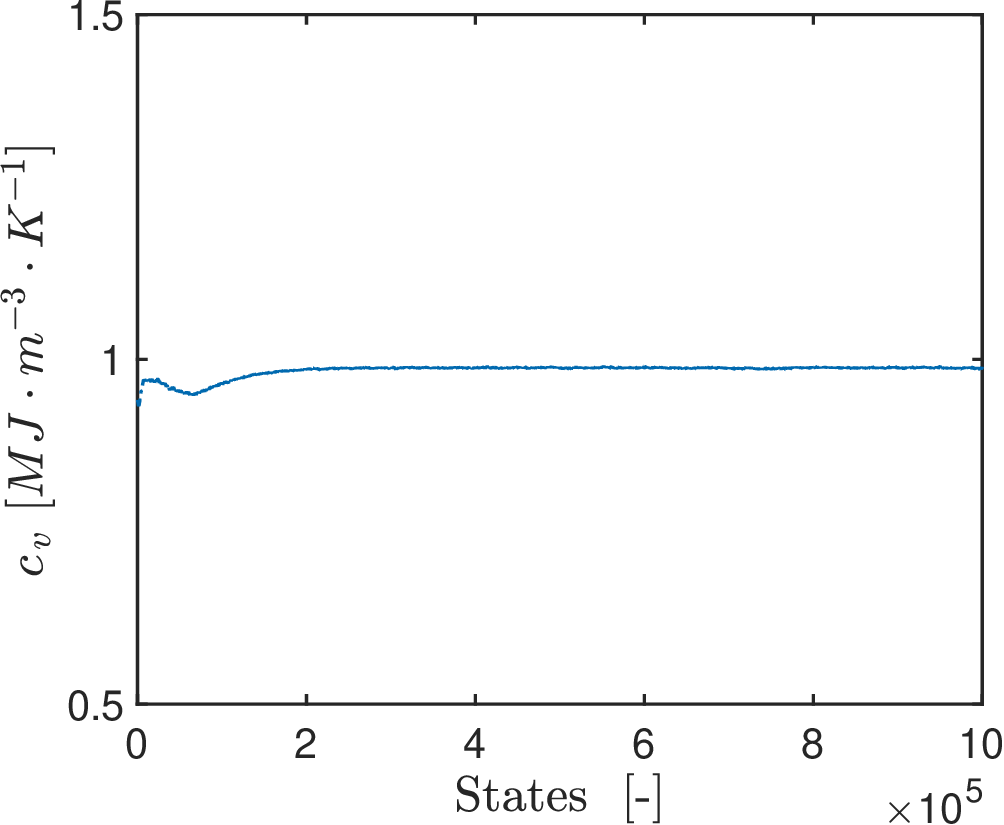}} \hspace{0.2cm}
\subfigure[\label{fig:k_c}]{\includegraphics[width=0.45\textwidth]{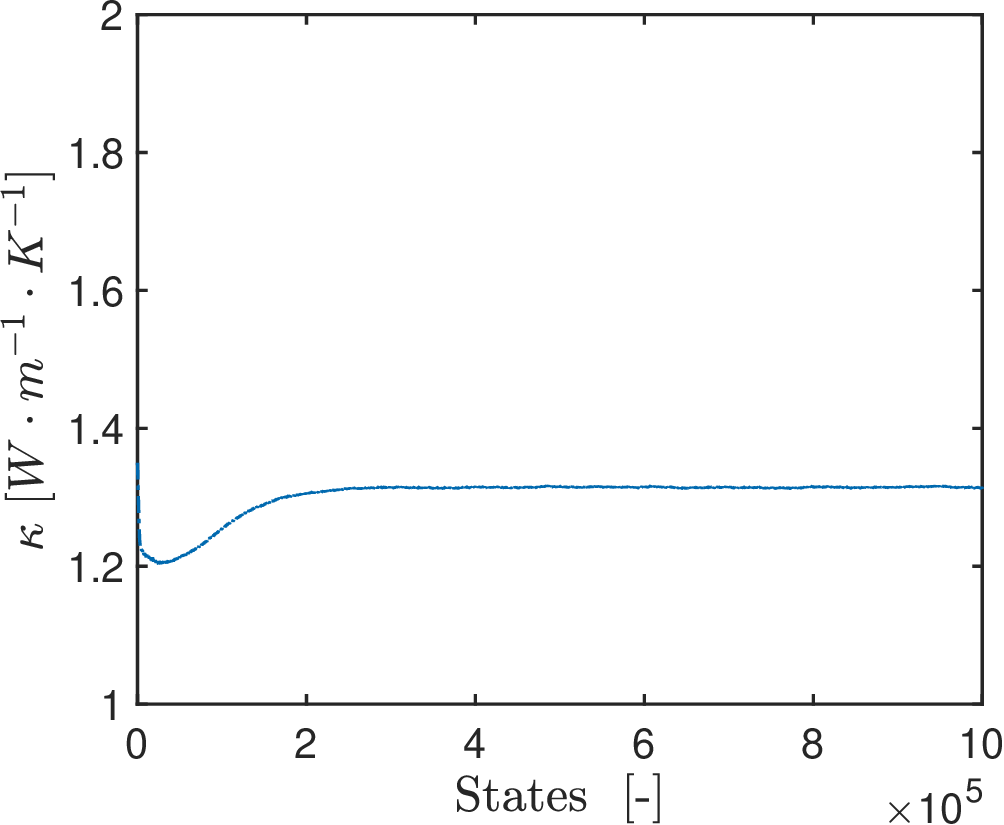}} \\
\subfigure[\label{fig:ht_c}]{\includegraphics[width=0.45\textwidth]{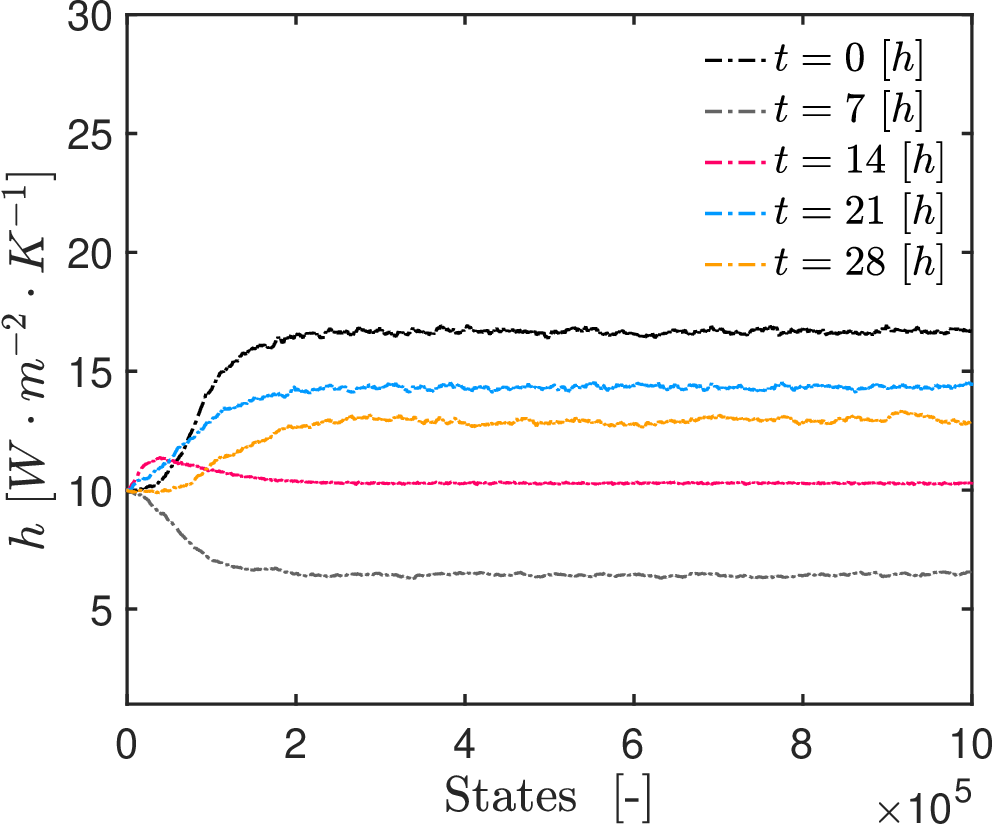}} \hspace{0.2cm}
\subfigure[\label{fig:ht_c_new_i}]{\includegraphics[width=0.45\textwidth]{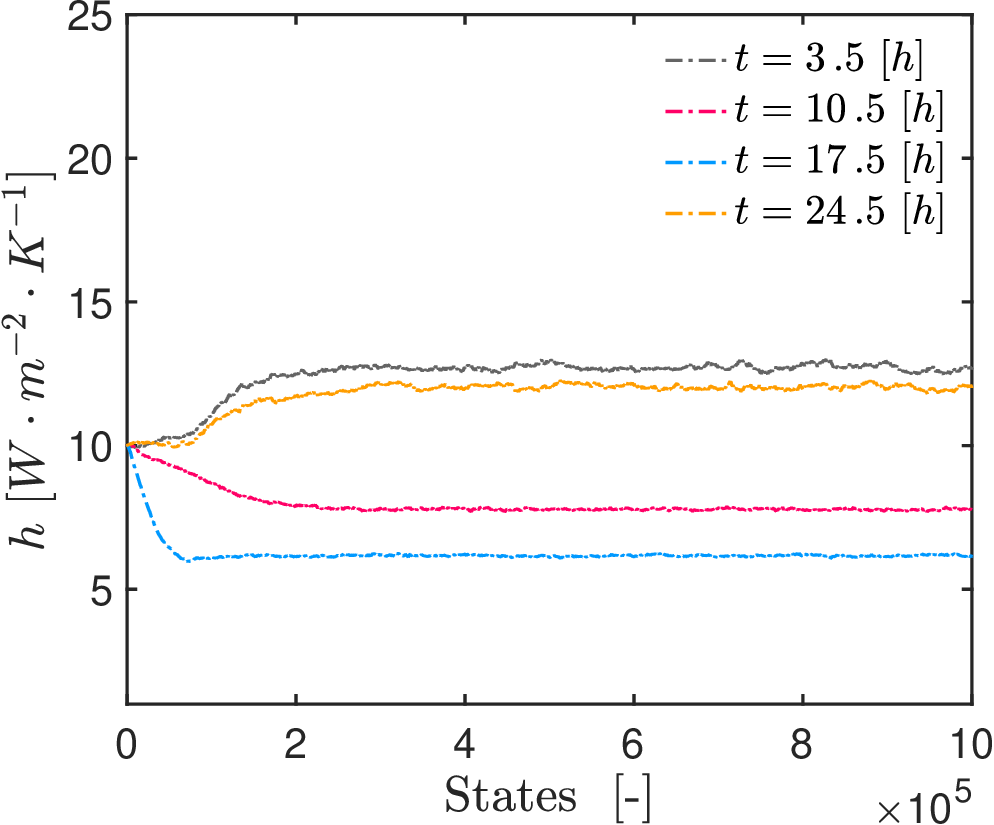}}
\\
\caption{Markov chains for $c$ \emph{(a)}, $\kappa$ \emph{(b)} and for various surface heat transfer coefficients \emph{(c)} - \emph{(d)}.}
\label{fig:markov_chains_c}
\end{center}
\end{figure}

\begin{figure}[!htb]
\begin{center}
{\includegraphics[width=0.45\textwidth]{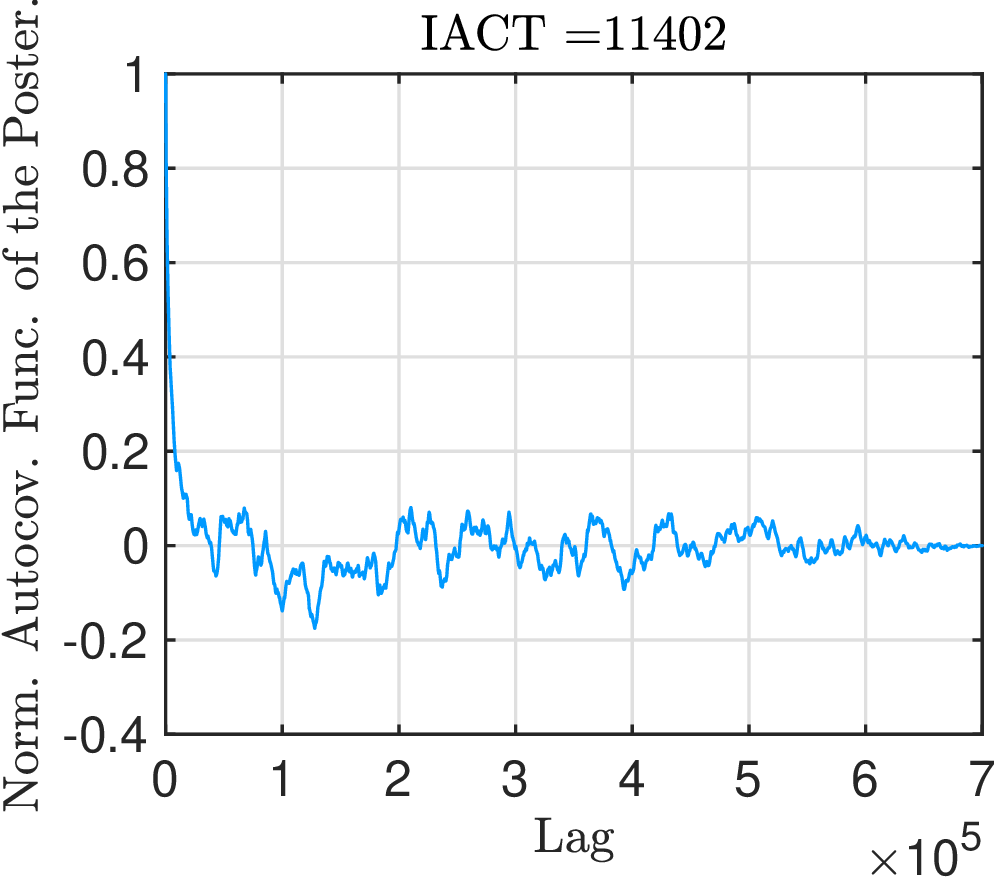}} 
\\
\caption{Variation of the norm of the autocovariance function of the posterior.}
\label{fig:autoc}
\end{center}
\end{figure}

From Figure \ref{fig:res_c} it can be seen that the acceptance rate is similar with Cases A and B, around $70 \%$. The evolution of the likelihood according to the number of state is given in Figure \ref{fig:posterior_c}. It shows that the chains converges to an equilibrium distribution after about $3 \cdot 10^{\,5}$ states. Figures \ref{fig:cv_c} - \ref{fig:ht_c_new_i} present the Markov chains for the parameters $c_v$, $\kappa$ and $h$ (at selected times), respectively. Such as Figure \ref{fig:posterior_c}, these figures show that the Markov chains, simulated with $10^{\,6}$ states, reached equilibrium after about $3 \cdot 10^{\,5}$ states. Consequently, such value is taken as burn-in period. Figure \ref{fig:autoc} represents the norm of the autocovariance function of the posterior , and the integrated autocorrelation time  (IACT). As expected, the norm of the autocovariance function of the posterior tends to zero remaining only influenced by a noise. However, we note that IACT was high, probably because of the large acceptance rate shown by Figure \ref{fig:res_c}.

\begin{figure}[!htb]
\begin{center}
\subfigure[\label{fig:c_hist_c}]{\includegraphics[width=0.45\textwidth]{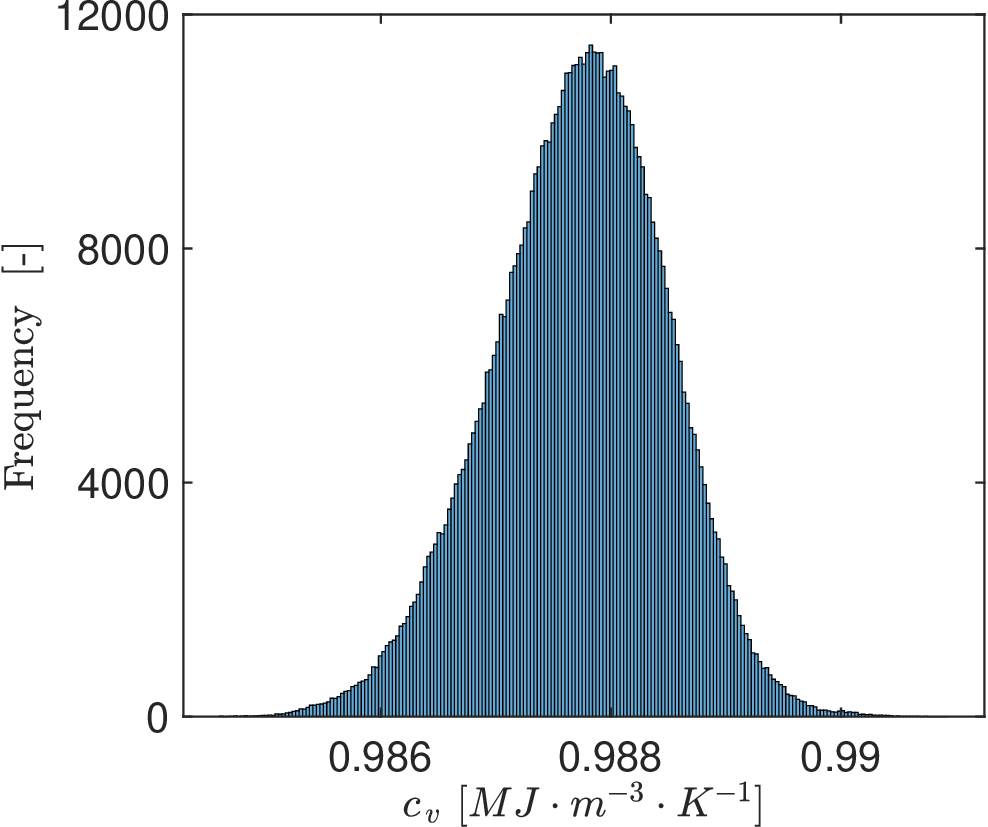}} \hspace{0.2cm}
\subfigure[\label{fig:k_hist_c}]{\includegraphics[width=0.45\textwidth]{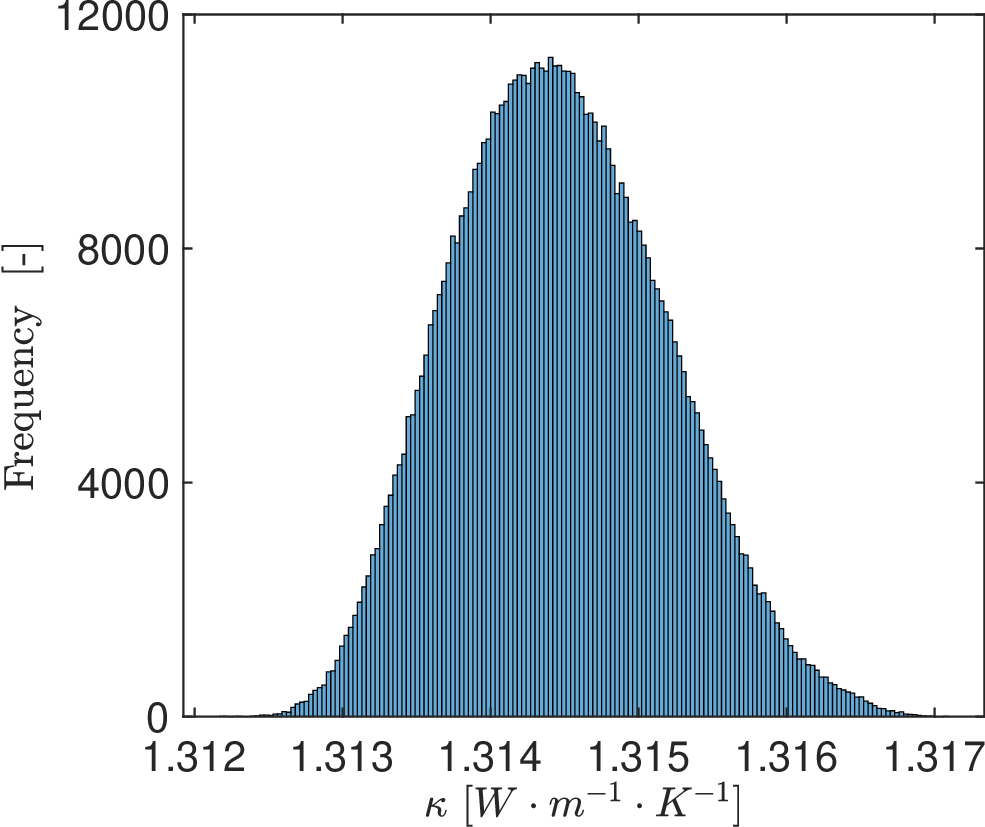}} \\
\subfigure[\label{fig:h2_c}]{\includegraphics[width=0.45\textwidth]{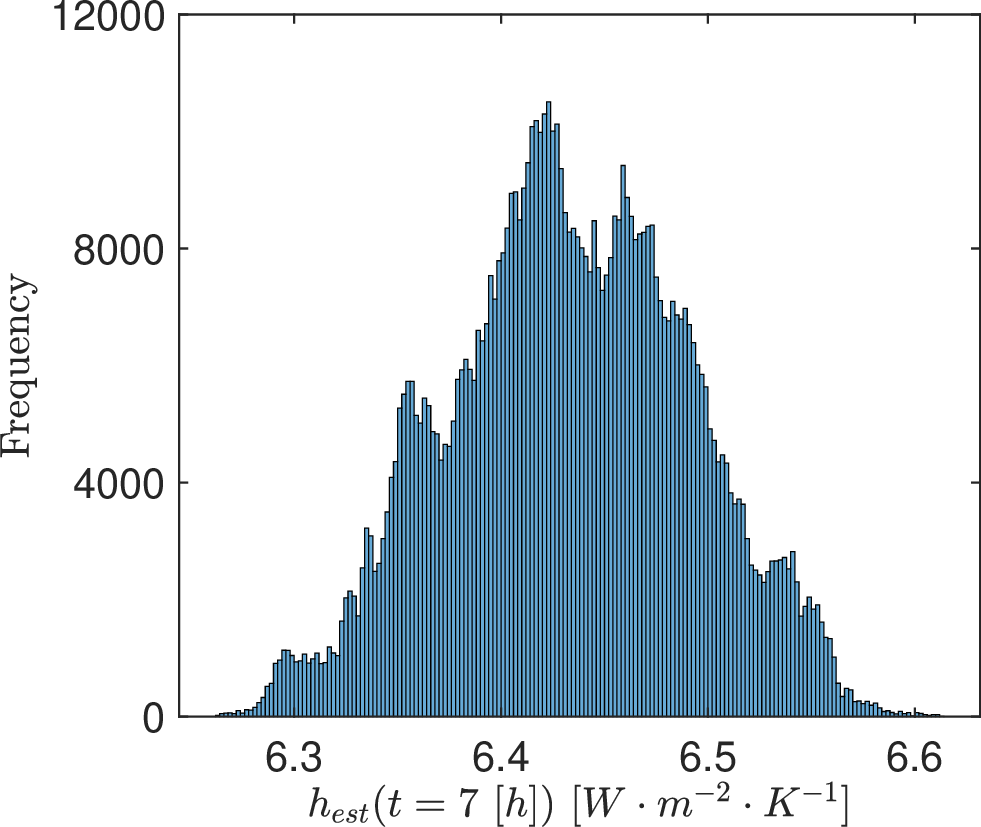}} \hspace{0.2cm}
\subfigure[\label{fig:h3_c}]{\includegraphics[width=0.45\textwidth]{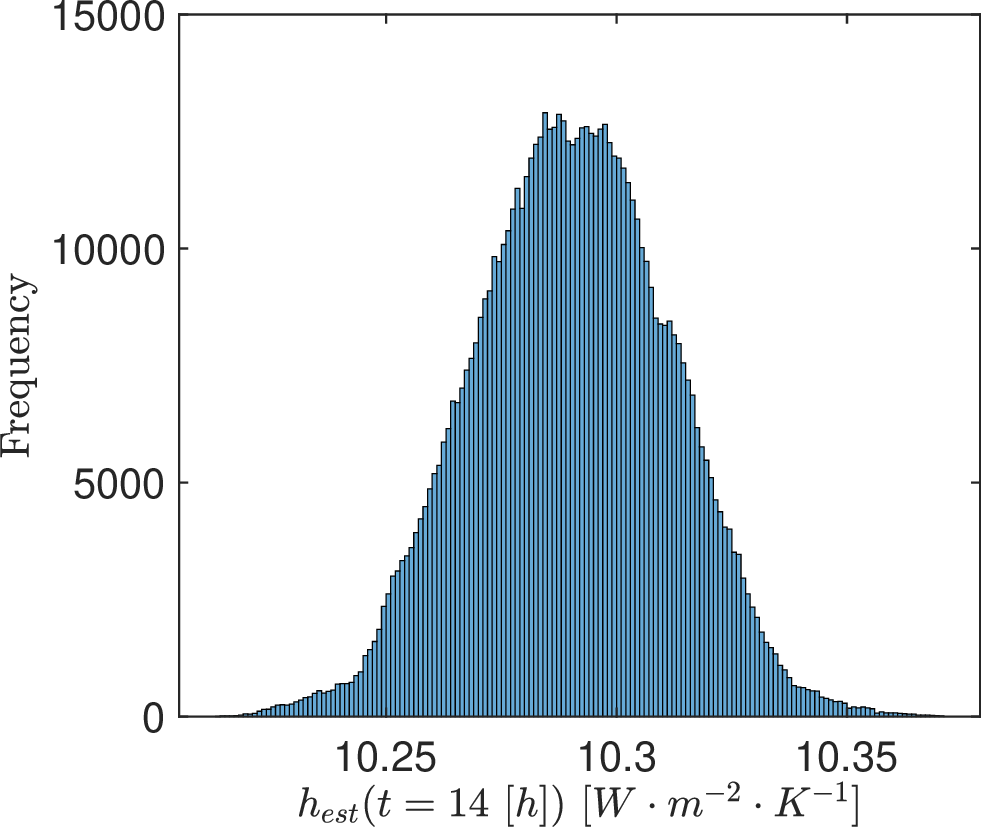}}  \\
\subfigure[\label{fig:h4_c}]{\includegraphics[width=0.45\textwidth]{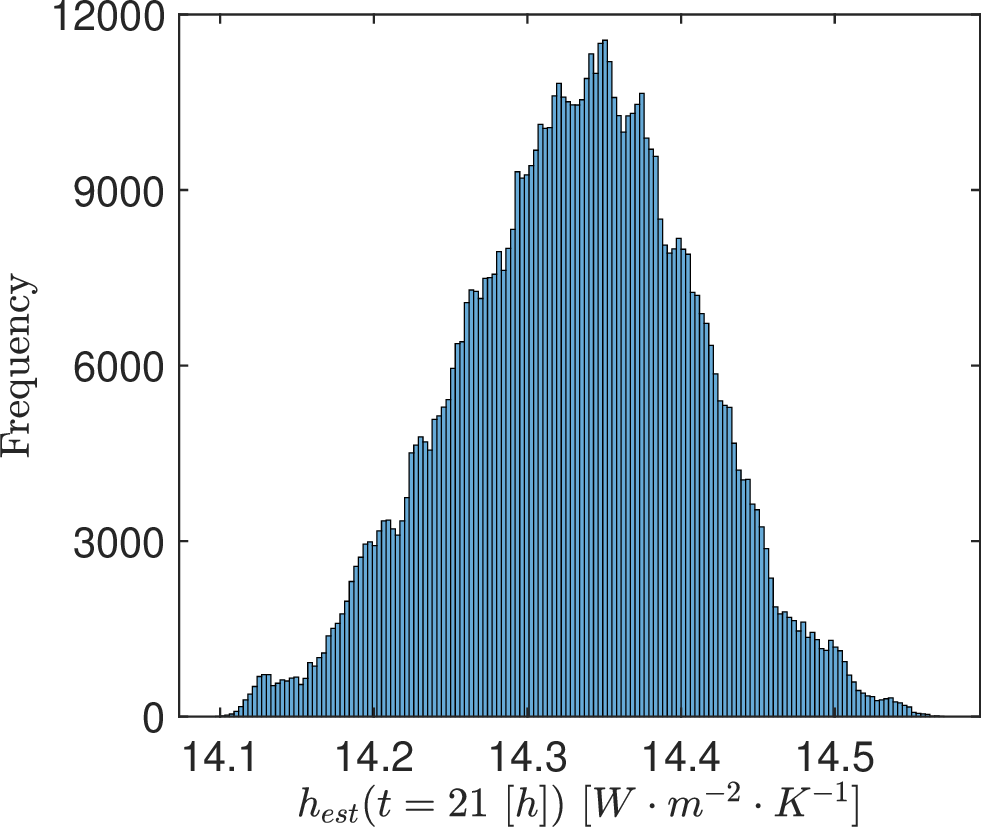}} \hspace{0.2cm}
\subfigure[\label{fig:h5_c}]{\includegraphics[width=0.45\textwidth]{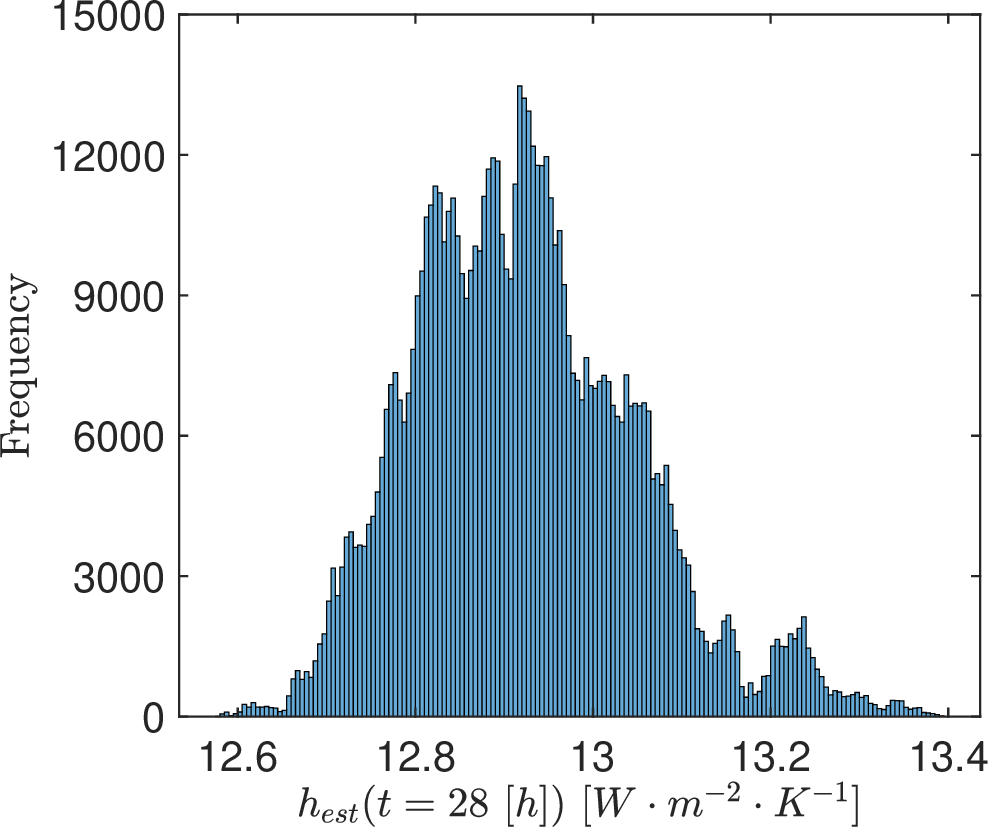}} \\
\caption{Histograms of the samples after the burn-in period for $c$ \emph{(a)}, for $\kappa$ \emph{(b)} and for $h$ at $t = 7 \, h$ \emph{(c)}, $t = 14 \, h$ \emph{(d)}, $t = 21 \, h$ \emph{(e)}, and $t = 28 \, h$ \emph{(f)}.}
\label{fig:c_histograms}
\end{center}
\end{figure}

The histograms of the samples of the Markov chains are presented in Figures \ref{fig:c_hist_c} - \ref{fig:h5_c}, for $C$, $\kappa$ and $h$ at selected times, respectively. From Figures \ref{fig:c_hist_c} and \ref{fig:k_hist_c}, the posterior distribution of the volumetric heat capacity and the thermal conductivity seems like Gaussian ones. Regarding to the surface heat transfer coefficient, Figures \ref{fig:h2_c} - \ref{fig:h5_c}, reveal that the marginal posteriors of the estimated surface heat transfer coefficients exhibited coefficients of variations smaller than $1 \, \%$. For time $t \, = 14 \, \ \mathsf{h}$, the histogram resembled a Gaussian distribution.

\begin{figure}[!htb]
\begin{center}
\subfigure[\label{fig:ht_mcmc_c}]{\includegraphics[width=0.45\textwidth]{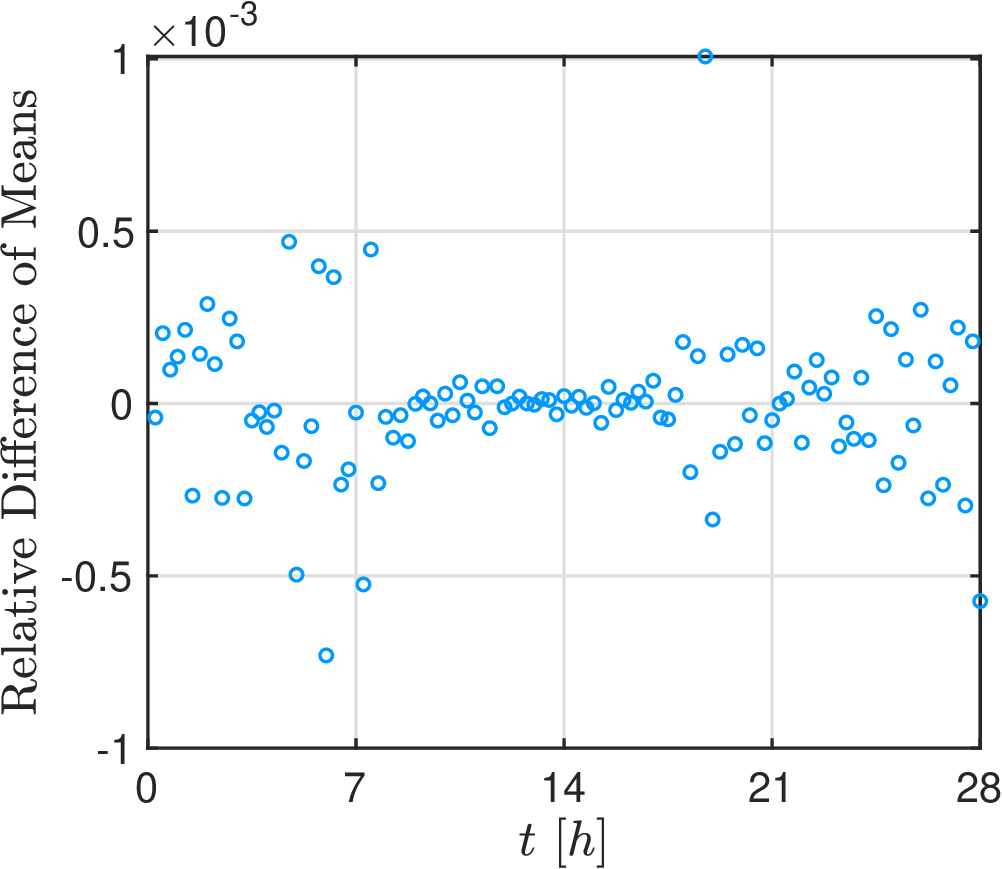}} \hspace{0.2cm}
\subfigure[\label{fig:h_i25}]{\includegraphics[width=0.45\textwidth]{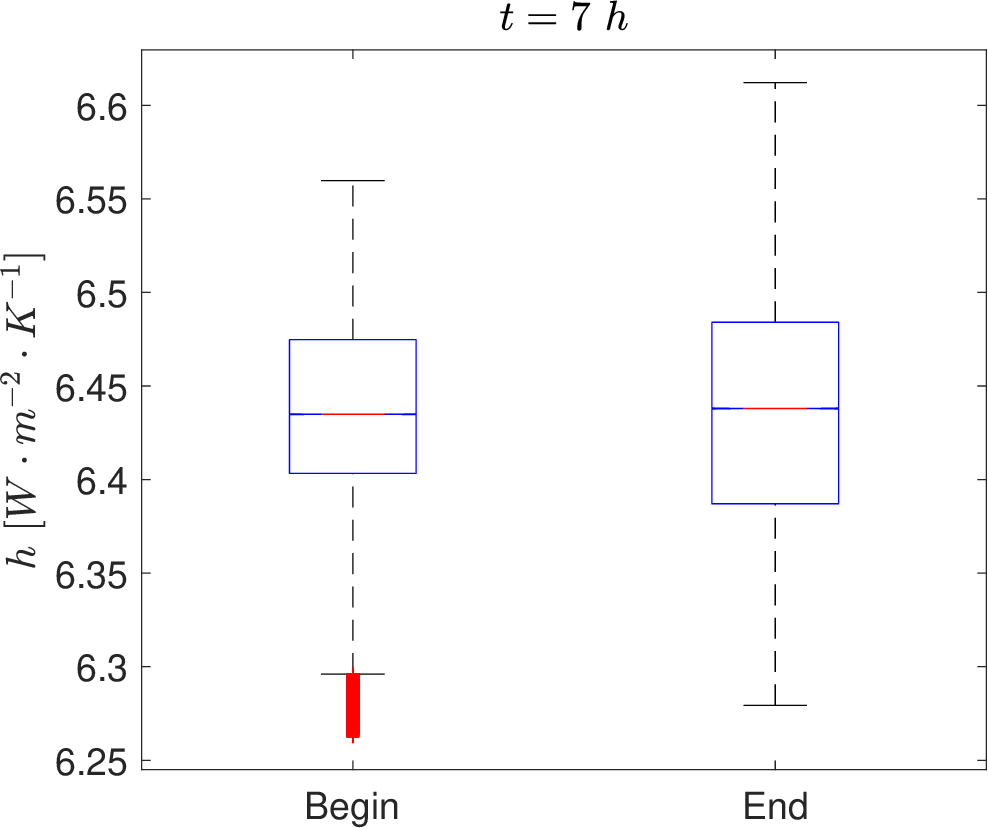}} 
\\
\subfigure[\label{fig:h_t14}]{\includegraphics[width=0.45\textwidth]{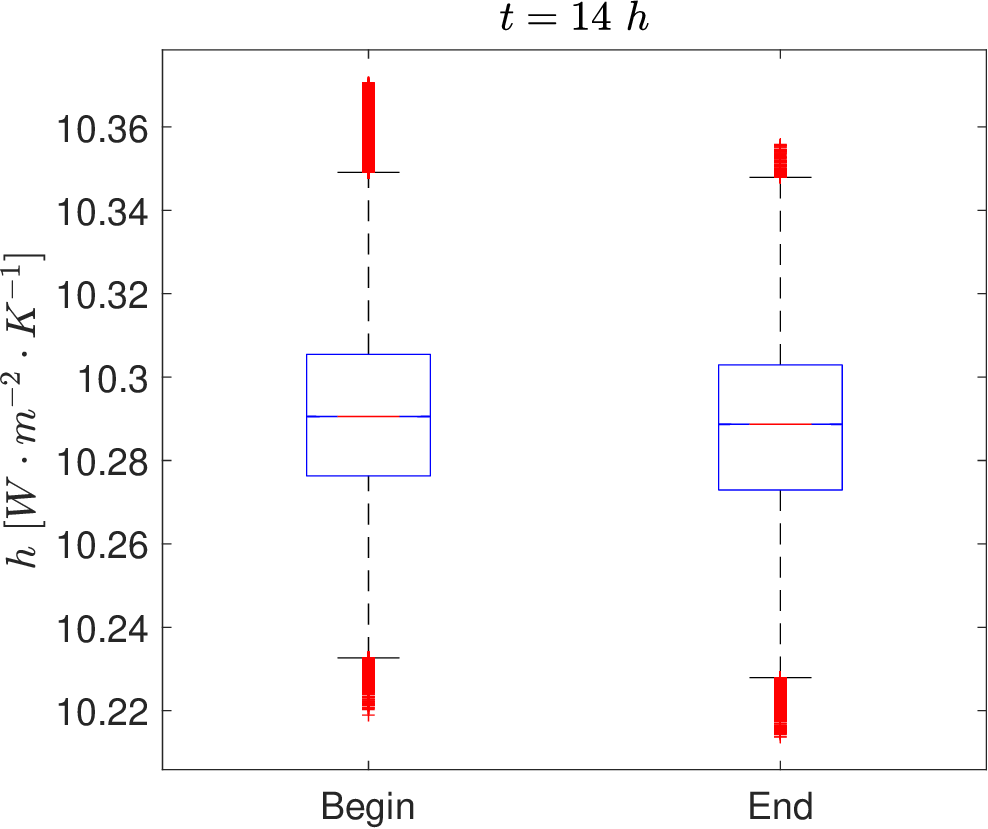}} \hspace{0.2cm}
\subfigure[\label{fig:h_i60}]{\includegraphics[width=0.45\textwidth]{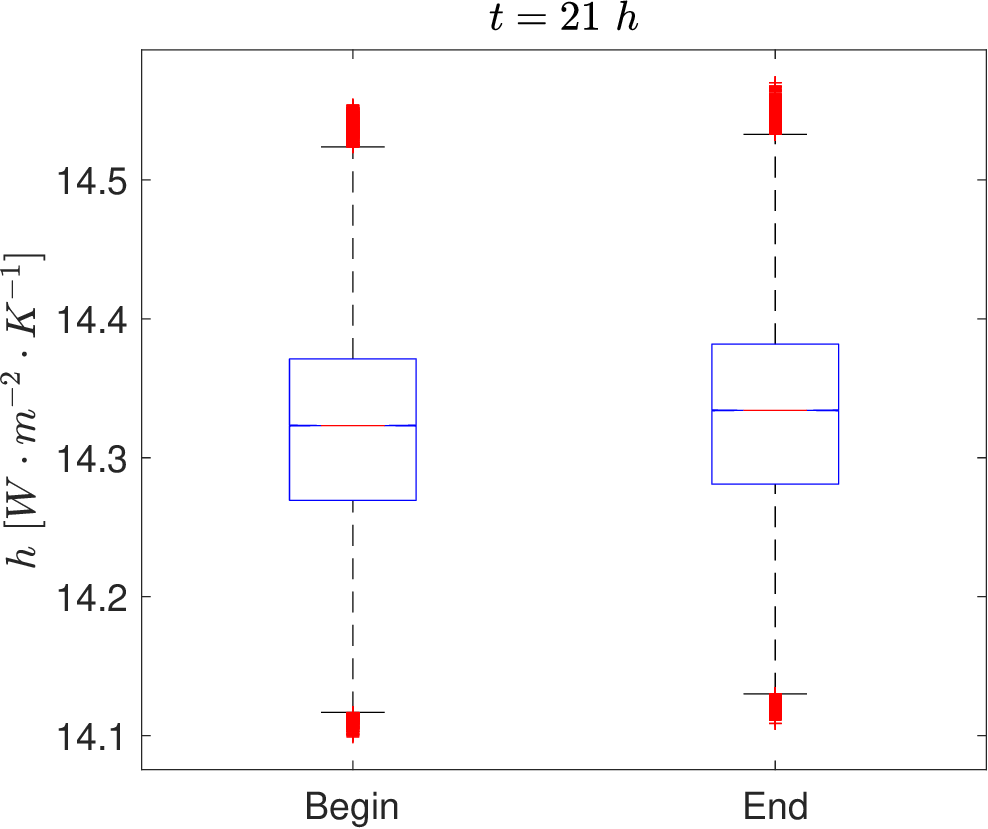}}
\\
\caption{Relative difference of means of the estimated surface heat transfer coefficients at each observation time \emph{(a)}. Range of the mean values of the surface heat transfer coefficient at $t = 7 \unit{h}$, \emph{(b)}  $t = 14 \unit{h}$ \emph{(c)} and $t = 21 \unit{h}$ \emph{(d)} between begin and last part of the Markov chain.}
\label{fig:Rel_dif}
\end{center}
\end{figure}

Those results are consistent with the analysis of the relative difference of means of the estimated surface heat transfer coefficients obtained with Geweke's analysis \cite{Geweke}, as illustrated in Figure \ref{fig:ht_mcmc_c}. The relative difference of the means computed with first $10 \, \%$ and last $50 \, \%$ of samples of the Markov chains after reaching equilibrium for each parameter \cite{Geweke}. The relative difference of means are higher for times between $1$ and $7 \ \mathsf{h}$, as well as between $17$ and $28 \ \mathsf{h}$. However, these differences are less than $10^{-3}$. The box plots of the samples of at the begin and end of the converged Markov chains are also shown by Figures \ref{fig:h_i25} - \ref{fig:h_i60}, for times $t = 7 \ \mathsf{h}$, $14 \ \mathsf{h}$ and  $21 \ \mathsf{h}$, respectively. Figures \ref{fig:h_i25} - \ref{fig:h_i60} show quite similar distributions at the beginning and end of the Markov chains. Figures \ref{fig:ht_mcmc_c} - \ref{fig:h_i60} show that the Markov chains reached equilibrium distributions with the burn-in period of $300 \ 000$ states considered for Case C.

The mean and standard deviation values of each unknown parameters, calculated after the burn-in period are presented in Table~\ref{tab:Initial guess_c}. 
 
\begin{table}[!htb]
\centering
\caption{Statistics for the marginal posterior distributions of the parameters.}
\label{tab:Initial guess_c}
\setlength{\extrarowheight}{.5em}
\begin{tabular}{|c|c|c|c|c|c|c|c|c|}
\hline
\hline 
\multicolumn{2}{|c|}{parameters} & \multicolumn{5}{c|} {$h(t)$} & $c_{\,v}$ &  $\kappa$ \\[5pt]
\hline 
\multicolumn{2}{|c|}{unit} & \multicolumn{5}{c|}{$\unit{\scriptstyle W\cdot m^{-2} \cdot K^{-1}}$} & $\unit{ \scriptstyle MJ \cdot m^{-3}\cdot K^{-1}}$ & $\unit{\scriptstyle W\cdot m^{-1} \cdot K^{-1}}$ \\[5pt]
\hline
\multicolumn{2}{|c|}{time instant} & $t = 0 \, h$ & $t = 7 \, h$ & $t = 14 \, h$ & $t = 21 \, h$ & $t = 28 \, h$ & -- &  -- \\[5pt]
\hline
\hline
estimated & mean & $16 \,.67$ & $6 \,.43$ & $10\,.29$ & $14\,.33$ & $12\,.92$ & $0 \,.99$ & $1 \,.31$\\
(burn-in $3 \cdot 10^5$) & std & $0 \,.09$ & $0 \,.06$ & $0 \,.02$ & $0 \,.08$ & $0 \,.13$ &  $0 \,. 0008$ & $0 \,. 0007$  \\
\hline
\hline
\hline
\end{tabular}
\end{table}

\begin{figure}
\begin{center} 
\vspace*{-2.0cm}
\subfigure[\label{fig:h_apr_est_c} $h(\,t\,)$]{\includegraphics[width=0.45\textwidth]{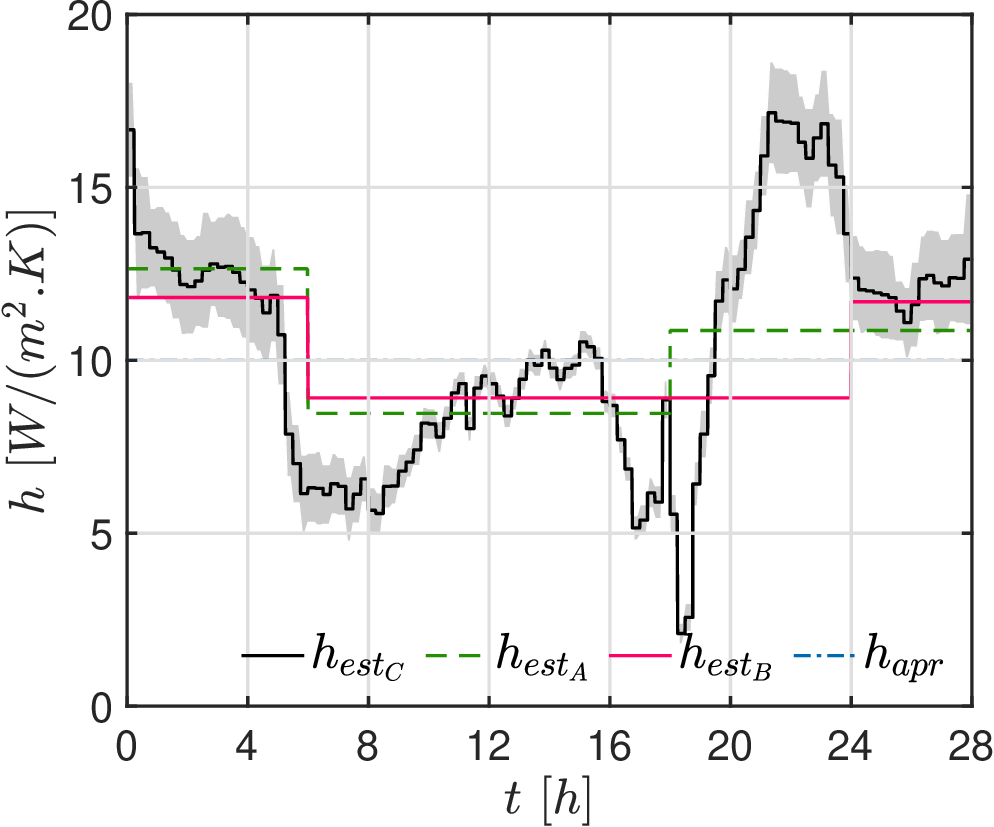}} \hspace{0.2cm}
\subfigure[\label{fig:ht_dif_gamma} $h(\,t\,)$ and $v(\,t\,)$]{\includegraphics[width=0.50\textwidth]{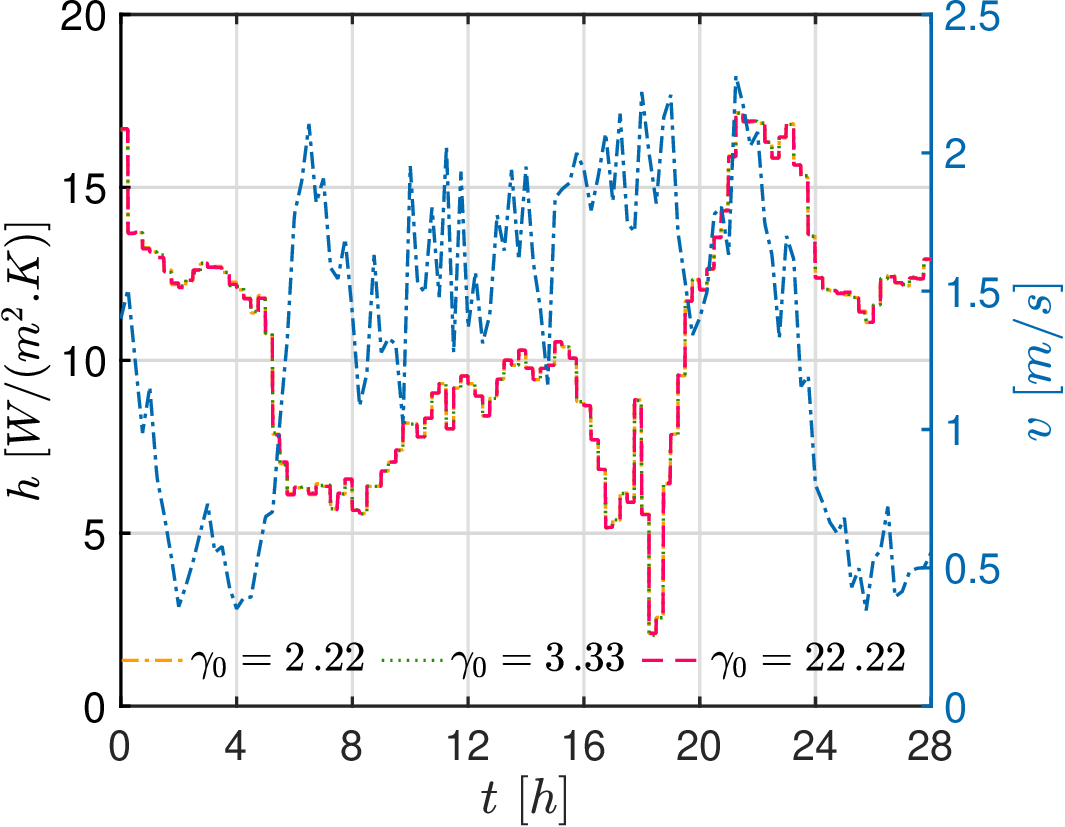}} 
\caption{Time varying surface heat transfer coefficient \emph{(a)}.Time varying surface heat transfer coefficient with different values of $\gamma_0$ and a wind velocity \emph{(a)}.}
\label{fig:ht_est_c}
\end{center}
\end{figure}

Figure~\ref{fig:h_apr_est_c} presents the surface heat transfer coefficients estimated with Cases A, B and C. The grey area for Case C gives the standard deviation of the estimated samples of the Markov chains. The behavior of the surface heat transfer coefficients of all cases are similar and consistent despite the different time discretizations used for each case. Figure~\ref{fig:ht_dif_gamma} shows the surface heat transfer coefficient estimated with Case C and the measured wind velocity. The wind velocity and surface heat transfer coefficient have similar behaviors. For the time periods $t \, \in \, \bigl[\, 0 \,,\, 6 \,\bigr] \ \mathsf{h}$ and $t \, \in \, \bigl[\, 21 \,,\, 28 \,\bigr] \ \mathsf{h}$ the surface heat transfer coefficient decreases with the velocity. For the period $t \, \in \, \bigl[\, 6 \,,\, 21 \,\bigr] \ \mathsf{h}\,$, the relation between both quantities is less obvious. It could be explained by several reasons. First, the velocity has fluctuations with high frequency and at different wind orientations as illustrated in Figure~\ref{fig:v}. The second reason is that the temperature and wind velocity measurements are carried out at different heights \cite{COHARD2018675}. 

Figure~\ref{fig:h_apr_est_c} shows that around $ t \egal 18 \ \mathsf{h}$ the surface heat transfer coefficient oscillates. It was examined if such oscillations were due to the ill-posed character of the inverse problem. Therefore, the inverse problem was solved for several \textit{a priori} values of the hyper parameter $\gamma_0 \in \bigl\{\, 2 \,. 22 \,,\, 3 \,. 33 \,, \, 22 \,. 22 \, \, \mathsf{m^{4} \cdot K^{2} \, \cdot W^{-2}} \bigr\}\,$. Results are presented in Figure~\ref{fig:ht_dif_gamma}. Different \textit{a priori} values of $\gamma_0$ provides consistent estimations of $h(\,t\,)\,$.

\begin{figure}[!htb]
\begin{center} 
\subfigure[\label{fig:Tx0_c} $x_m = 0 \ \mathsf{cm}$]{\includegraphics[width=0.45\textwidth]{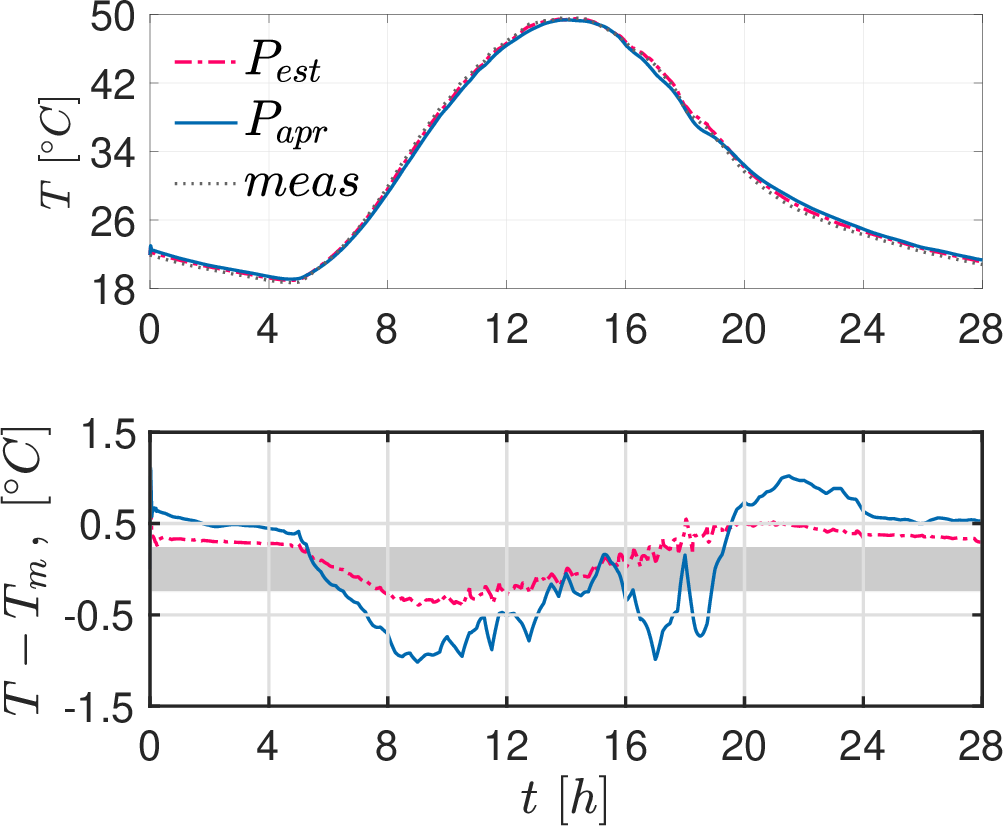}} \hspace{0.2cm}
\subfigure[\label{fig:Tx1_c} $x_m = 1 \ \mathsf{cm}$]{\includegraphics[width=0.45\textwidth]{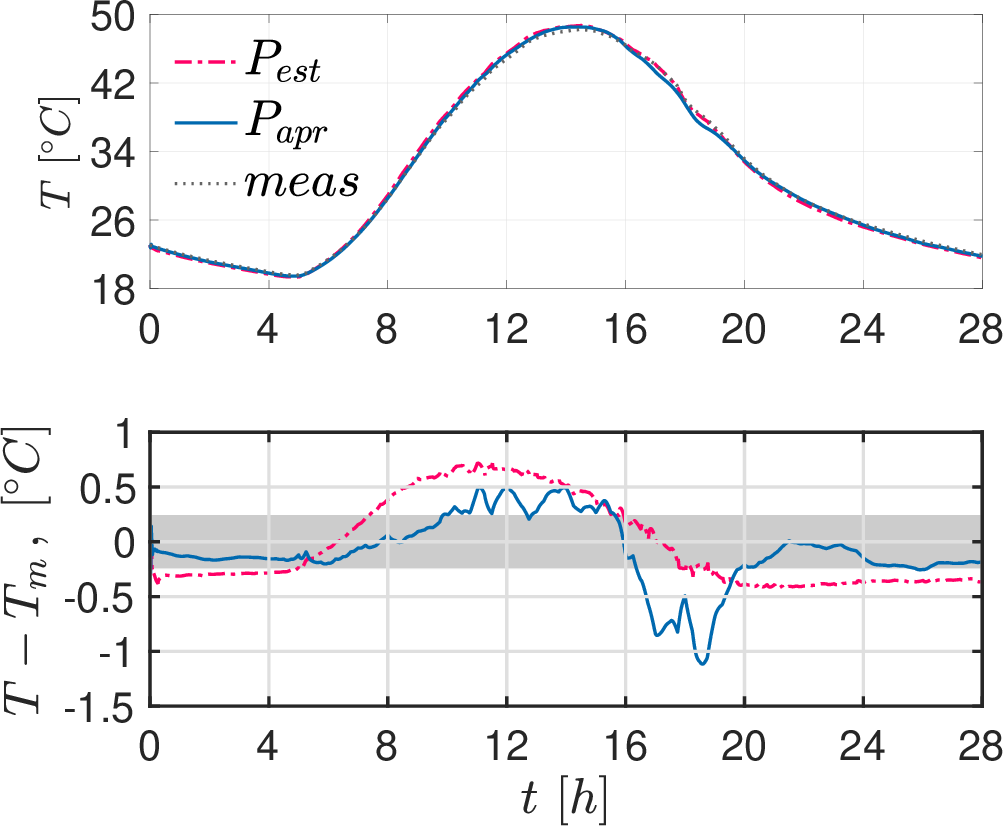}} \\
\subfigure[\label{fig:Tx2_c} $x_m = 2 \ \mathsf{cm}$]{\includegraphics[width=0.45\textwidth]{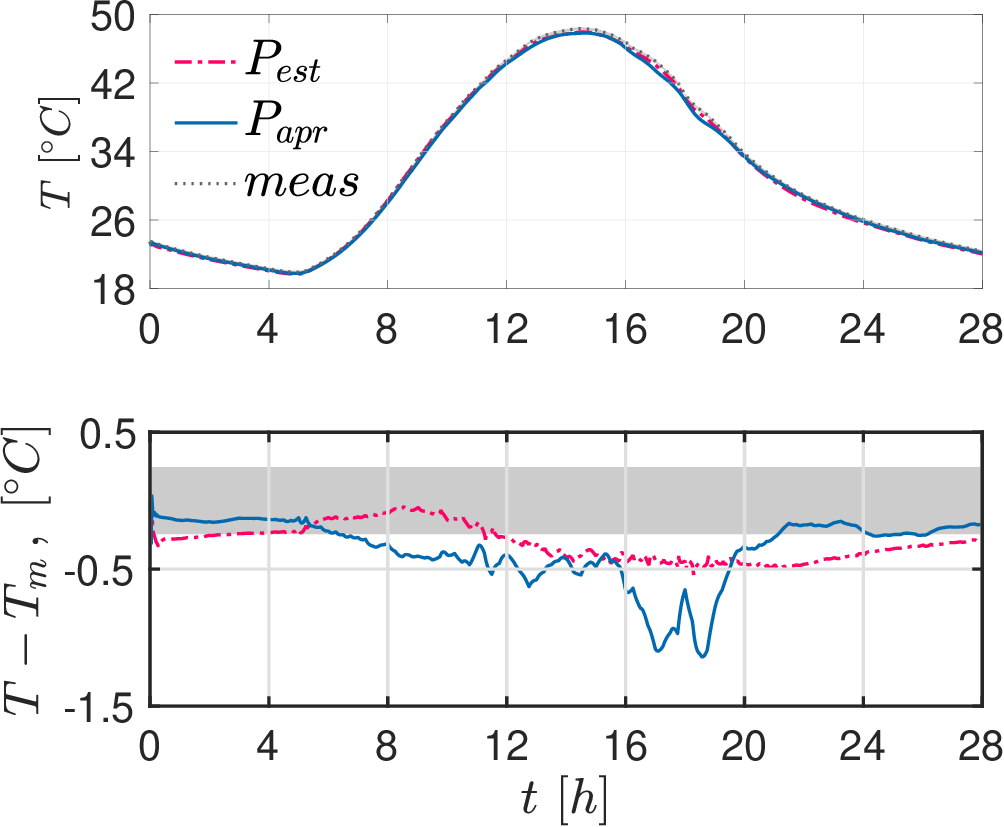}} \hspace{0.2cm}
\subfigure[\label{fig:Tx3} $x_m = 3 \ \mathsf{cm}$]{\includegraphics[width=0.45\textwidth]{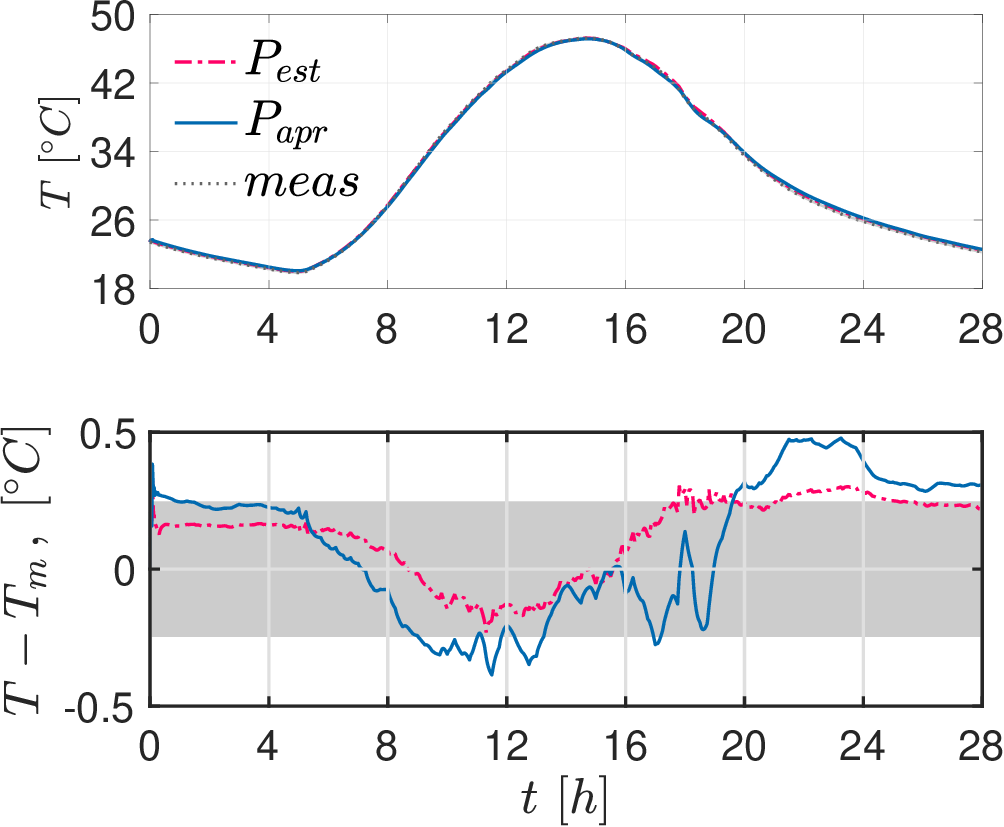}} \\
\subfigure[\label{fig:Tx4_c} $x_m = 4 \ \mathsf{cm}$]{\includegraphics[width=0.45\textwidth]{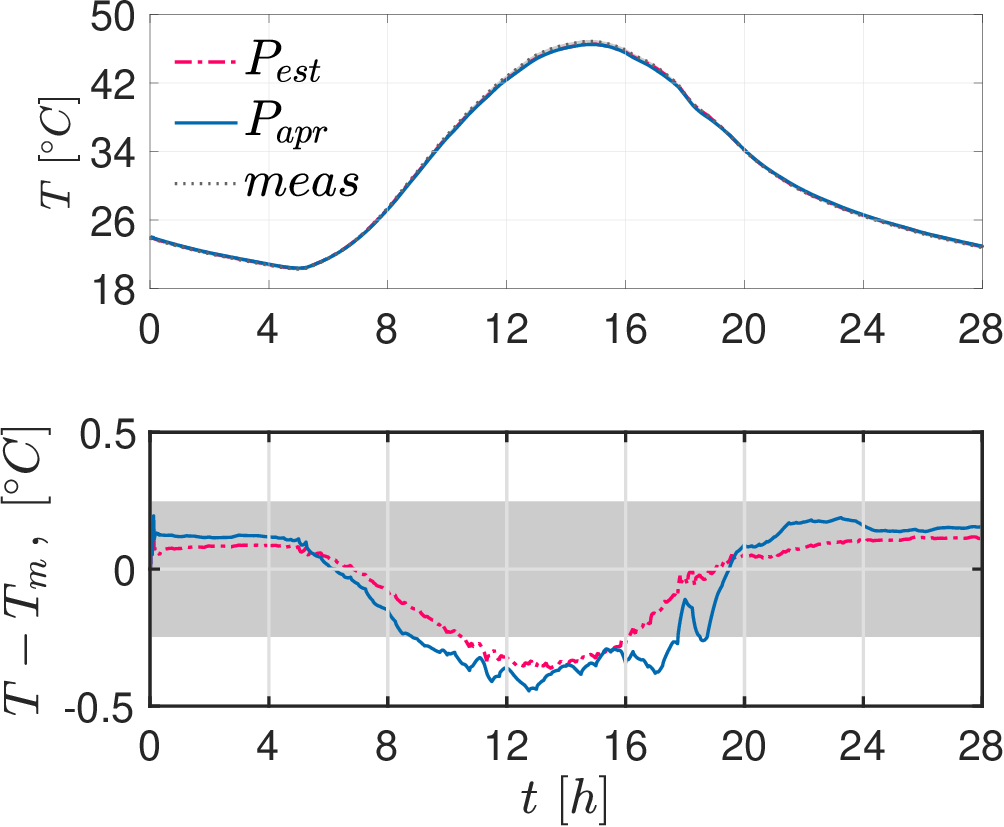}}\hspace{0.2cm}
\caption{ Computed temperatures with mean of the estimations and literature values, and measured values of temperature at several sensor locations \emph{(a)} - \emph{(e)}.}
\label{fig:DP_results_c}
\end{center}
\end{figure}

After the estimation procedure, the direct problem is computed with the mean values of the posterior distributions of each parameter, as well as with the literature values of the parameters. The corresponding temperatures computed at five measurement positions and the respective measurements are compared in Figures~\ref{fig:Tx0_c}-\ref{fig:Tx4_c}. The residuals are also presented in those Figures. Temperatures computed with the estimated values are in better accordance with the measurements than ones computed with the literature values of the parameters. Moreover, residuals of the Case C are lower and less correlated than those for Cases A and B. The agreement is improved by at least $8 \, \%$ with the time discretization of the surface heat transfer coefficient used in Case C, as compared to Cases A and B.

\begin{figure}[!htb]
\begin{center}
{\includegraphics[width=0.45\textwidth]{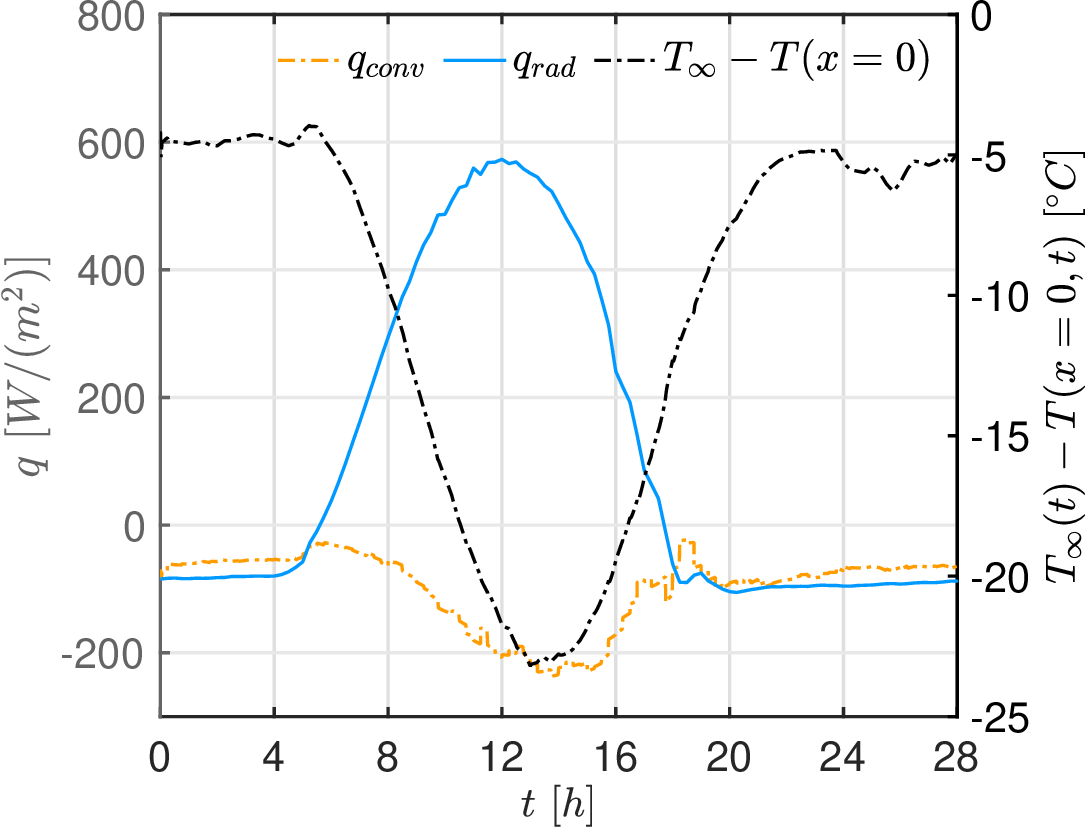}} 
\\
\caption{Time varying heat fluxes at the surface of the ground and temperature difference $T_{\infty}(t) - T( x = 0, t )$}
\label{fig:q_c_r}
\end{center}
\end{figure}


Effects of the computed convection and measured radiation heat fluxes were also examined. These fluxes are presented in Figure \ref{fig:q_c_r}. This figure also presents the temperature difference $T_{\infty}(t)  - T(x=0,t)$ given by the black line. Around $18 \ \mathsf{h}\,$, the radiation heat flux decreases and becomes negative probably because of the decrease in the solar irradiation. Meanwhile, during the whole experiment the convection flux is negative, since the ground surface is warmer than the ambient air. As a result of the energy balance at the ground surface, the temperature increases from $6$ to $14 \ \mathsf{h}\,$, and then it decreases afterwards shown by Figure~\ref{fig:Tx0_c}. The instability on the surface heat transfer coefficient can be due to the radiation heat flux that changed sign at $18 \ \mathsf{h}\,$, which might have affected the estimation procedure.  

A small oscillation of the radiation flux is observed at  $t \egal 18 \ \mathsf{h}\,$ in Figure~\ref{fig:q_c_r}. To check the influence of such oscillation on the estimation of the surface heat transfer coefficient, the inverse problem has been solved considering a filtered radiation heat flux, but there were not modifications on the estimated surface heat transfer coefficient. The radiation heat flux was smoothed using a moving average filter of $1 \,. 5 \, \mathsf{h}$ .

All those analysis lead to conclude that estimation of surface heat transfer coefficient were performed with success and lead to a better agreement between the model predictions and the measured temperatures.

\section{Influence of the surface heat transfer coefficient on the energy balance in a urban scene}
\label{sec:urban_scale}

Previous section estimated the time varying surface heat transfer coefficient at the interface between the ground and the outside air. In the context of better assessing the energy balance of the radiation flux in the urban environment, a simulation is performed with a micro-climatic simulation tool \cite{Azam_2025}. The time step is set to $15 \ \mathsf{min}$ with $8 \ \mathsf{days}$ of run-off simulation.

\begin{figure}[!htb]
\begin{center}
{\includegraphics[width=0.7\textwidth]{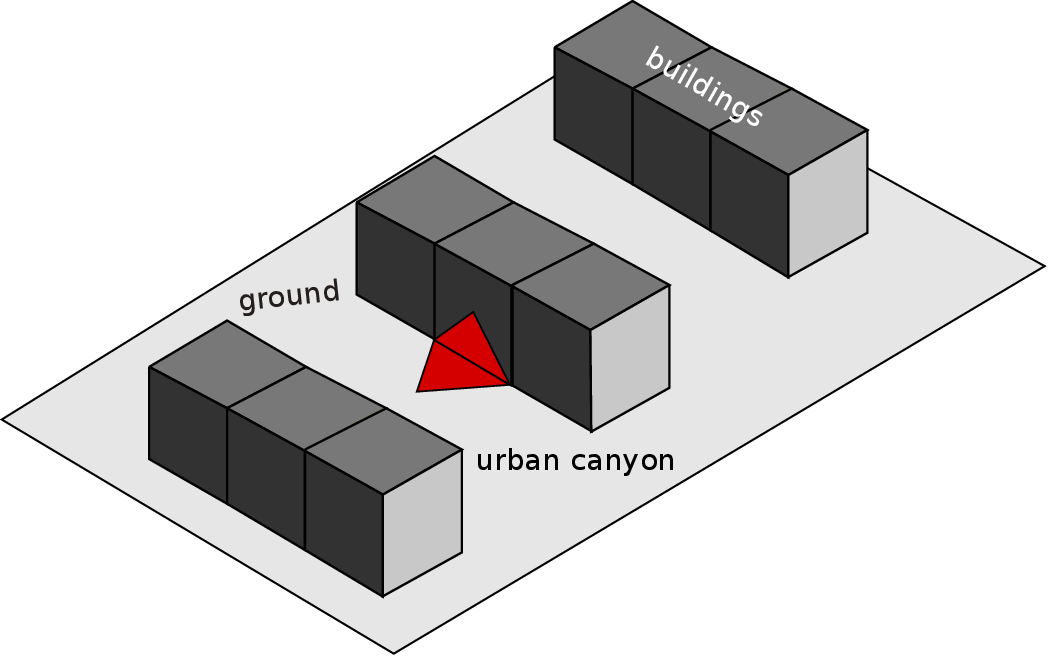}}
\caption{Illustration of the urban scene with the zone of interests highlighted in red.}
\label{fig:urban_scene}
\end{center}
\end{figure}

The geometric configuration represents a real urban environment, inspired from the experimental platform \cite{Djedjig}. A schematic illustration of the case study is presented in Figure~\ref{fig:urban_scene}. The platform is composed of $9$ empty concrete tanks arranged in three rows, representing a reduced version of buildings and streets. The streets are set with a ratio building height / street width of $1.1$ and oriented North-South. Further details on the geometrical configurations can be obtained in the original work \cite{Djedjig}. The tanks are composed of concrete wall of $4 \ \mathsf{cm}$ which thermal properties are: $\kappa \egal 2.4 \ \mathsf{W\cdot m^{-1} \cdot K^{-1}}$ and $C \egal 1.9 \ \mathsf{MJ \cdot m^{-3}\cdot K^{-1}}\,$  \cite{Djedjig}. For the first $5 \ \mathsf{cm}$, the thermophysical properties of the soil are the one estimated in Case C and reported in Table~\ref{tab:Initial guess_c}. For the deeper layers, the thermal properties are the one considered in \cite{AZAM2018728}. The measured temperature at $z \egal 75 \ \mathsf{cm}$ is used as boundary condition in the ground. The solar reflectivity (albedo) of the wall is $0.64$ and $0.36$ for the ground \cite{AZAM2018728}. The short-wave radiation flux is decomposed as direct and diffuse component calculated using the global component measured during the experimental campaign (see Section~\ref{ssec:description}) and with the method described in \cite{ERBS1982293}. The outside temperature is directly given from the experimental data. Last, the long-wave radiation flux is computed from the measurement of incoming long-wave (thermal infrared) radiation. 

\begin{figure}[!htb]
\begin{center} 
\subfigure[\label{fig:urban_h_ft}]{\includegraphics[width=0.45\textwidth]{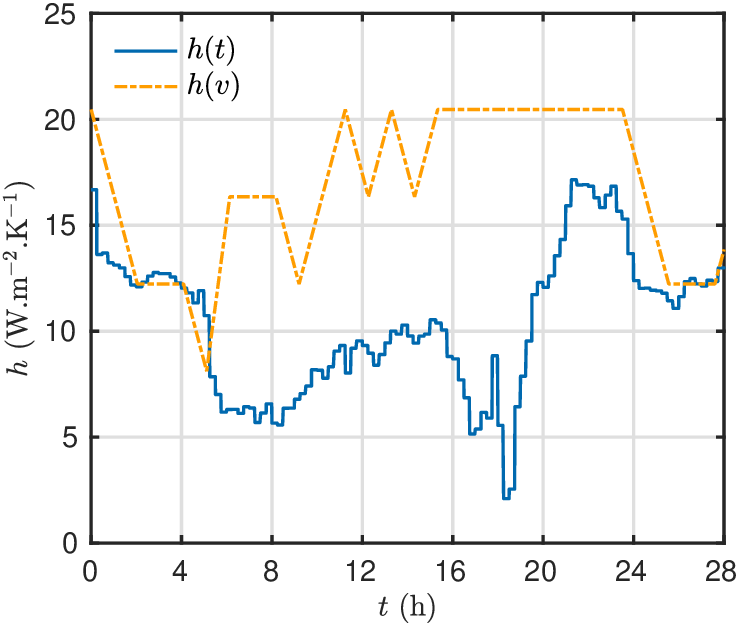}} \hspace{0.2cm}
\subfigure[\label{fig:urban_Tsurf}]{\includegraphics[width=0.45\textwidth]{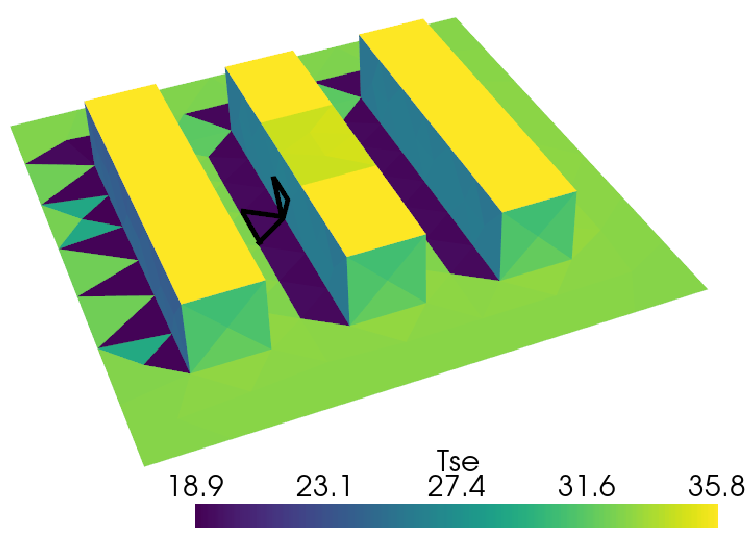}} \\
\subfigure[\label{fig:urban_Tg} ground]{\includegraphics[width=0.45\textwidth]{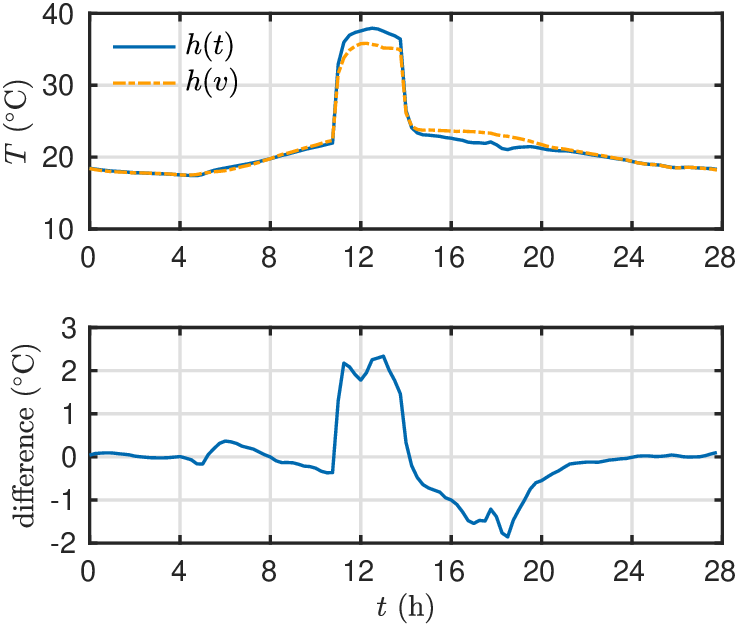}} \hspace{0.2cm}
\subfigure[\label{fig:urban_Tw} wall]{\includegraphics[width=0.45\textwidth]{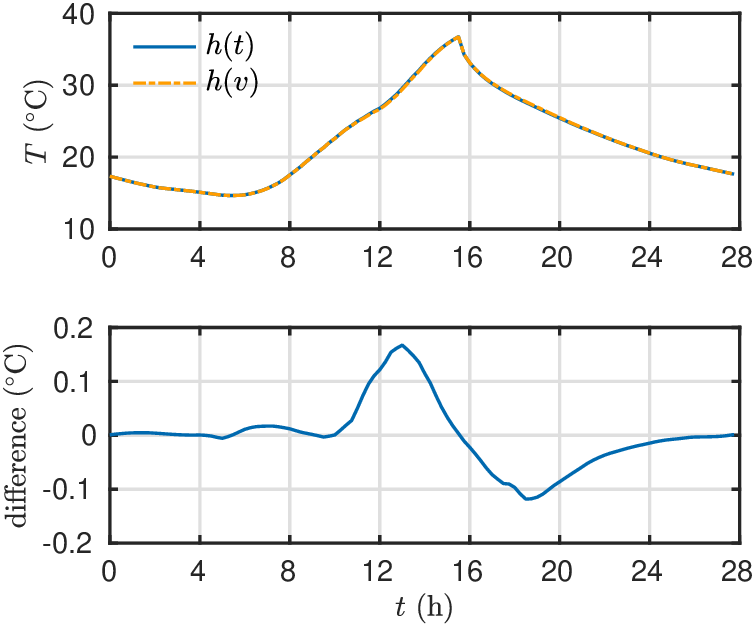}} \\
\subfigure[\label{fig:urban_GLOg} ground]{\includegraphics[width=0.45\textwidth]{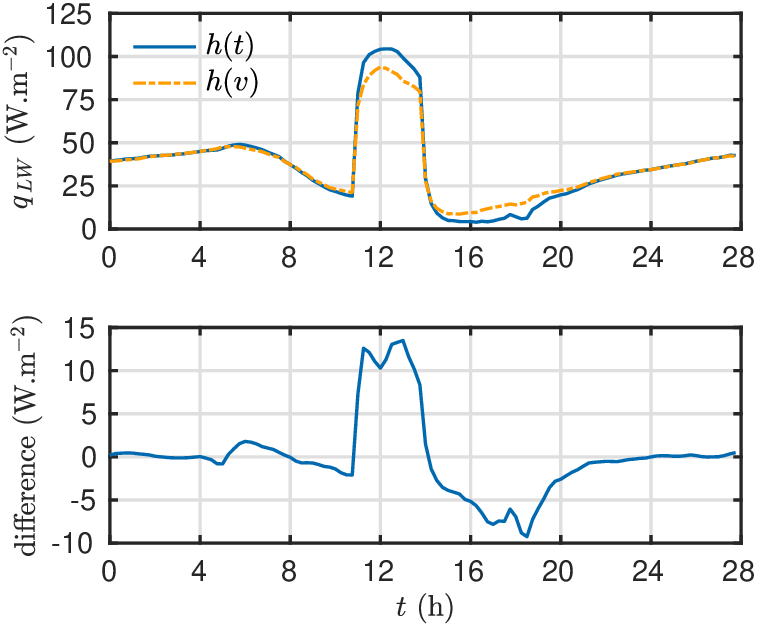}} \hspace{0.2cm}
\subfigure[\label{fig:urban_GLOw} wall]{\includegraphics[width=0.45\textwidth]{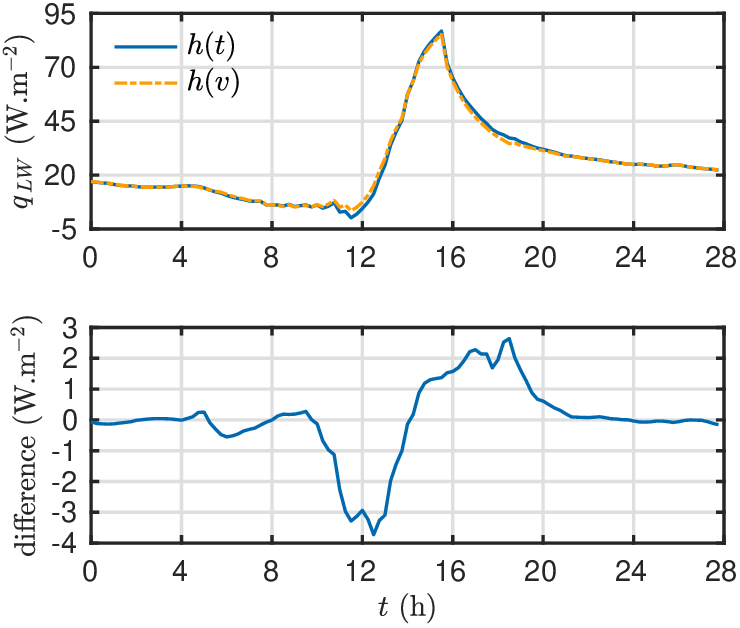}} 
\caption{Time varying ground surface coefficient \emph{(a)} for the urban environment simulation. Computed surface temperature at $t \egal 10.75 \ \mathsf{h}$ for the case $h(\,t\,)$ \emph{(b)}. Time varying surface temperature and long wave radiation flux for the ground \emph{(c,e)} and the building wall \emph{(d,f)}.}
\label{fig:urban_results}
\end{center}
\end{figure}

Two simulations are performed, considering the whole long-wave radiation balance between the surfaces and the heat conduction through the ground and walls for the all urban scene. Results are presented for two zones of interest highlighted in red in Figure~\ref{fig:urban_scene}. The first simulation corresponds to standard simulation: the ground surface heat transfer coefficient is computed using the wind air velocity and the following empirical correlations \cite{fr_std}: 
\begin{align*}
    h(\,v\,) \egal 4 \plus 4 \cdot v \,.
\end{align*}
In most of the standard simulations at district or building scale, to save computational effort with CFD simulations, the wind velocity is taken from the nearest airport meteorological station. In this case, it corresponds to the Nantes-Atlantique airport (around $10 \ \mathsf{km}$ distant from the case study site), obtained from \cite{infoclimat} for the exact day of investigations (June 6$^{th}$, 2004). The second simulations is carried out using the time varying ground surface heat transfer coefficient retrieved in case C. The coefficients considered for the two simulations are presented in Figure~\ref{fig:urban_h_ft}. Figure~\ref{fig:urban_Tsurf} shows the surface temperature of the whole urban environment computed using $h(\,t\,)\,$. The temperature is lower in the urban canyon due to the shading effects. The difference between the two simulations are given in Figures~\ref{fig:urban_Tg}--\ref{fig:urban_GLOw}. The zones of interests for the analyze of thermal balance are highlighted in red in Figure~\ref{fig:urban_scene} (and black Figure~\ref{fig:urban_Tsurf}), corresponding to the surface ground in the canyon and a wall of a close building. A difference of more than $2 \ \mathsf{\degC}$ is observed on the surface ground temperature. The difference is lower for the surface building wall temperature with a magnitude of $0.1 \ \mathsf{\degC}$ around midday. Regarding the received long-wave radiation flux $q_{\,LW} \ \unit{W\cdot m^{\,-2}}\,$, it increases during the day with the raise of the surface temperature of the urban scene. The difference between the two simulations reaches almost $15 \ \mathsf{W\cdot m^{\,-2}}$ and $4 \ \mathsf{W\cdot m^{\,-2}}$ for the ground and building wall, respectively. This analysis highlights the requirement of modeling accurately the surface heat transfer coefficient for reliable assessment of the energy balance at the urban scale. 

\section{Conclusion}

In this work, an inverse one-dimensional heat conduction problem is solved to estimate the unknown thermal conductivity, volumetric heat capacity and time varying surface heat transfer coefficients. The Bayesian framework with the MCMC algorithm is used to explore the posterior distributions of the unknown parameters based on their prior distributions and on the likelihood function that models the measurement errors. The measured temperature are obtained from an experimental campaign with \emph{in-situ} monitoring of ground with $5$ sensors. Three approximations are considered for the time varying surface heat transfer coefficients. For the first two case studies, A and B, the coefficient is defined as piece-wise continuous for three time periods. Gaussian priors are applied for the unknown parameters. Retrieved parameters lead to a lower residual of the temperature, compared to the direct computations using parameters from the literature. The third case study, case C, estimates a value of the surface heat transfer coefficient for each time interval when measurements are carried out (each $15 \ \mathsf{min}$). For this parameter, a Gaussian smoothness prior is considered. Results show a more accurate prediction of the calibrated model than for Cases A and B.

Last, a comparison at the urban scale is performed between a standard simulation, which assumes a ground surface coefficient computed with the wind velocity measured at the nearest airport meteorological station, and a simulation using the estimated time varying coefficient from case C. Results highlight the importance of accurate modeling of this parameter for reliable assessment of the energy balance of the urban environment.

\newpage~

\section*{Nomenclature}

 \begin{tabular}[!htbp]{|cll|}
 \hline
 \multicolumn{3}{|c|}{\emph{Latin letters}} \\
 $C$ & volumetric heat capacity & $[\mathsf{J \cdot m^{-3} \cdot K^{\,-1})}]$ \\
 $c_{\,\rho}$ & specific heat capacity & $[\mathsf{J \cdot kg^{\,-1}\cdot K^{\,-1})}]$ \\
 $L$ & ground depth & $[\mathsf{m}]$ \\
 $t_{\,f}$ & final time & $[\mathsf{s}]$ \\
 $T$ & temperature & $[\mathsf{K}]$ \\
 $T_{\infty}$ & temperature of the air & $[\mathsf{K}]$ \\
 $T_{g}$ & temperature of the ground & $[\mathsf{K}]$ \\
  $T_{in}$ & initial temperature & $[\mathsf{K}]$ \\
 $h$ & surface heat transfer coefficient & $[\mathsf{W \cdot m^{\,-2}\cdot K^{\,-1}}]$ \\
 $q_{\infty}\,,\,q_{\,LW}$ & radiation flux & $[\mathsf{W \cdot m^{\,-2}}]$ \\
 $t$ & time  & $[\mathsf{s}]$ \\
 $x$ & space coordinate & $[\mathsf{m}]$ \\
 $U$ & dimensionless temperature & $[\, - \,]$ \\
 $k$ & dimensionless heat conductivity & $[\, - \,]$ \\
 $Fo$ & Fourier number & $[\, - \,]$ \\
 $Bi^L$ & Biot number & $[\, - \,]$ \\
 \hline
 \end{tabular}

\bigskip

 \begin{tabular}[!htbp]{|cll|}
 \hline
 \multicolumn{3}{|c|}{\emph{Greek letters}} \\
 $\kappa$ & thermal conductivity & $[\mathsf{W \cdot m^{\,-1}\cdot K^{\,-1}}]$ \\
 $\alpha$ & thermal diffusivity & $[\mathsf{m^{\,2} \cdot s^{\,-1}}]$ \\
 $\rho $ & density & $[\mathsf{kg \cdot m^{\,-3}}]$ \\
 \hline
 \end{tabular}

\bigskip

 \begin{tabular}[!htbp]{|cl|}
 \hline
 \multicolumn{2}{|c|}{\emph{Subscripts and superscripts}} \\
 $*$ & dimensionless parameter \\
 $apr$ & \textit{a priori} parameter value \\
 $est$ & estimated parameter value \\
 \hline
 \end{tabular}
 
 \section*{Acknowledgments}

The authors acknowledge DSG 2021 and the French National Research Agency (ANR) as part of the ``JCJC CE-22 AAPG 2023'' program (project TOPS) for the financial support. This research is funded by the Science Committee of the Ministry of Education and Science of the Republic of Kazakhstan (Grant No. AP19175447). 

\bibliographystyle{unsrt}
\bibliography{references.bib}

 \appendix

\end{document}